\documentclass[11pt,a4paper]{article}
\usepackage{jheppub}

\usepackage{amsmath}
\usepackage{verbatim}
\usepackage{graphicx}
\usepackage{mathrsfs}
\usepackage{appendix}
\usepackage{caption}
\usepackage{float}
\usepackage{subfig}
\usepackage{mathtools}\usepackage{mathtools}
\usepackage{braket}

\def\cJ{\mathcal{J}}
\def\cL{\mathcal{L}}

\def\cN{\mathcal{N}}

\def\cP{\mathcal{P}}

\def\cH{\mathcal{H}}

\def\bL{\overline{\mathcal{L}}}

\def\bea#1\ena{\begin{align}#1\end{align}}
\def\nn{\nonumber\\}

\def\pd{\partial}

\def\a{\alpha}
\def\ta{\tilde\alpha}

\def\e{\epsilon}

\def\k{\kappa}

\def\w{\omega}

\def\dett[#1,#2,#3,#4]{\left|%
\begin{array}{cc} #1 & #2 \\ #3 & #4 \end{array} \right|}
\def\matt[#1,#2,#3,#4]{\left(%
\begin{array}{cc} #1 & #2 \\ #3 & #4 \end{array} \right)}
\def\mattt[#1,#2,#3,#4,#5,#6,#7,#8,#9]{\left(%
\begin{array}{ccc} #1 & #2 & #3 \\ #4 & #5 & #6 \\ #7 & #8 & #9 \end{array} \right)}
\def\dettt[#1,#2,#3,#4,#5,#6,#7,#8,#9]{\left|%
\begin{array}{ccc} #1 & #2 & #3 \\ #4 & #5 & #6 \\ #7 & #8 & #9 \end{array} \right|}
\def\vect[#1,#2]{\left(%
\begin{array}{cc} #1 \\ #2  \end{array} \right)}
\def\vectt[#1,#2]{\left(%
\begin{array}{cc} #1 \\ #2  \end{array} \right)}
\def\vecttt[#1,#2,#3]{\left(%
\begin{array}{cc} #1 \\ #2 \\ #3 \end{array} \right)}
\def\sqvect[#1,#2]{\left[%
\begin{array}{cc} #1 \\ #2  \end{array} \right]}
\def\wbvect[#1,#2]{\left\{%
\begin{array}{cc} #1 \\ #2  \end{array} \right\}}
\def\tvect[#1,#2]{\left(%
\begin{array}{cc} #1 & #2  \end{array} \right)}
\def\tvectt[#1,#2]{\left(%
\begin{array}{cc} #1 & #2  \end{array} \right)}
\def\tvecttt[#1,#2,#3]{\left(%
\begin{array}{ccc} #1 & #2 & #3 \end{array} \right)}
\def\ket[#1]{\left| #1 \right\rangle}
\def\bra[#1]{\left\langle #1 \right|}
\def\brak[#1,#2]{\left\langle #1 | #2 \right\rangle}
\def\braket[#1,#2]{\left\langle#1\right.\hspace{-2.pt}\left| #2\right\rangle}
\def\pair[#1,#2]{\left\langle #1 , #2 \right\rangle}
\def\be{\begin{equation}}
\def\ee{\end{equation}}
\def\ben{\begin{equation*}}
\def\een{\end{equation*}}

\usepackage{mathtools}

\newcommand{\DeclareAutoPairedDelimiter}[3]{%
	\expandafter\DeclarePairedDelimiter\csname Auto\string#1\endcsname{#2}{#3}%
	\begingroup\edef\x{\endgroup
		\noexpand\DeclareRobustCommand{\noexpand#1}{%
			\expandafter\noexpand\csname Auto\string#1\endcsname*}}%
	\x}
\DeclareAutoPairedDelimiter{\p}{(}{)}

\title{Beyond Wilson? Carroll from current deformations}

\author[a]{Arjun Bagchi,} 
\author[b]{Aritra Banerjee,} 
\author[a]{Saikat Mondal,}
\author[c]{Debangshu Mukherjee,} 
\author[d]{and Hisayoshi Muraki.}
\author[]{\\} 
 
\affiliation[a]{Indian Institute of Technology Kanpur, Kalyanpur, Kanpur 208016. INDIA}
\affiliation[b]{Birla Institute of Technology and Science, Pilani Campus, Pilani, Rajasthan 333031, INDIA. }
\affiliation[c]{Asia Pacific Center for Theoretical Physics, POSTECH, Pohang 37673, KOREA.}
\affiliation[d]{Center for Geometry and Physics, Institute for Basic Science (IBS), Pohang 37673, KOREA. \\}

\emailAdd{abagchi@iitk.ac.in,aritra.banerjee@pilani.bits-pilani.ac.in,
saikatmd@iitk.ac.in,debangshu.mukherjee@apctp.org,hmuraki@ibs.re.kr }

\abstract{At extreme energies, both low and high, the spacetime symmetries of relativistic quantum field theories (QFTs) are expected to change with Galilean symmetries emerging in the very low energy domain and, as we will argue, Carrollian symmetries appearing at very high energies. The formulation of Wilsonian renormalisation group seems inadequate for handling these changes of the underlying Poincare symmetry of QFTs and it seems unlikely that these drastic changes can be seen within the realms of relativistic QFT. We show that contrary to this expectation, changes in the spacetime algebra occurs at the very edges of parameter space. 
In particular, we focus on the very high energy sector and show how bilinears of $U(1)$ currents added to a two dimensional (massless) scalar field theory deform the relativistic spacetime conformal algebra to conformal Carroll as the effective coupling of the deformation is dialed to infinity. We demonstrate this using both a symmetric and an antisymmetric current-current deformation for theories with multiple scalar fields. These two operators generate distinct kinds of quantum flows in the coupling space, the symmetric driven by Bogoliubov transformations and the antisymmetric by spectral flows, both leading to Carrollian CFTs at the end of the flow.}

\preprint{}

\begin{document}
\maketitle


\section{Introduction}
The modern understanding of quantum field theory is to a large part based on the wonderful picture of renormalisation of Wilson \cite{RevModPhys.47.773}, where high energy modes are systematically integrated out in order to reach an effective low energy theory. Given a quantum field theory, this process of continually integrating out modes takes the theory from an arbitrary point on parameter space to an infra-red (IR) fixed point, governed by a conformal field theory (CFT). There is also an ultra-violet (UV) fixed point governed by a UV CFT. 

\medskip

Deformations of a quantum field theory can be relevant, marginal and irrelevant. CFTs are fixed points in these renormalisation group (RG) flows and the parameter space may contain CFTs other than UV and IR CFTs. Relevant deformations move one along RG trajectories from the UV CFT to the IR CFT through intermediate points which are a series of quantum field theories. Marginal deformations (the exactly marginal ones) govern flows along a line of CFTs while irrelevant deformations have the promise of running up RG flows at the cost of locality in the resulting QFTs. These are notoriously difficult to track and work on irrelevant deformations has been limited before the recent advances in a specific class of deformations called the $T{\bar T}$ deformations. See \cite{Jiang:2019epa} for a review and a comprehensive list of papers on the subject. When there are other conserved charges in the theories, one can also have other solvable deformations e.g.  $J {\bar T}$ deformations \cite{Guica:2017lia}. 

\medskip

The space of all QFTs can be constructed as RG flows away from these fixed points governed by CFTs. The very ambitious programme of trying to classify all quantum field theories can be recast into the question of classifying all CFTs, which is perhaps a bit more tractable. One of the major goals of the conformal bootstrap programme \footnote{See, for example, \cite{Poland:2018epd} for an exposition to the subject and detailed list of references.} is to attempt answering this question. 

\medskip

\subsection{Possible breakdown of Wilsonian RG at very low energies} 
The formulation of Wilsonian RG for relativistic quantum field theories does not allow changes in the spacetime algebra. However, it is very easy to envision a scenario where this should run into trouble. If we are interested in low energies and real world phenomena in e.g. condensed matter physics, the non-relativistic approximation is often a very good approximation. So, formulated generally, a renormalisation group flow should be able to reduce a theory which is a relativistic QFT to a Galilean QFT if we are sufficiently low in energies. The framework of RG that we use seems to be incapable to understanding changes in the spacetime algebra that are required in order to go from the relativistic to the non-relativistic world {\footnote{Wilson's original formulation \cite{RevModPhys.47.773} of course was done in the context of non-relativistic field theories. But even here the theory cannot accommodate changes in the spacetime algebra. Moving up the RG flow to a spacetime algebra which is Poincare instead of Galilean cannot be handled.}}. The various attempts at non-relativistic effective field theories in the literature suffer from this inherent and basic problem. These theories are not Galilean invariant theories, but Poincare invariant theories masquerading as non-relativistic field theories. 

\medskip

This paper aims to initiate a programme devoted to the understanding of the very low and the very high energy sectors of relativistic theories where the spacetime algebras themselves get transformed. We have discussed above how one can move down the RG flow lines from higher to lower energies via relevant deformations and when we go to sufficiently low energies we should end up in effect in the non-relativistic world where all the relativistic degrees of freedom have been integrated out and the theory should become Galilean invariant. We also stressed that this is possibly not achievable in the usual Wilsonian framework, since there is no notion of a change in spacetime algebra in this regard. We now discuss the ultra-high energy regime which would be the focus of the current paper. 

\subsection{Emergence of Carroll at very high energies} 
\begin{figure}[t]
	\centering
\includegraphics[width=6cm]{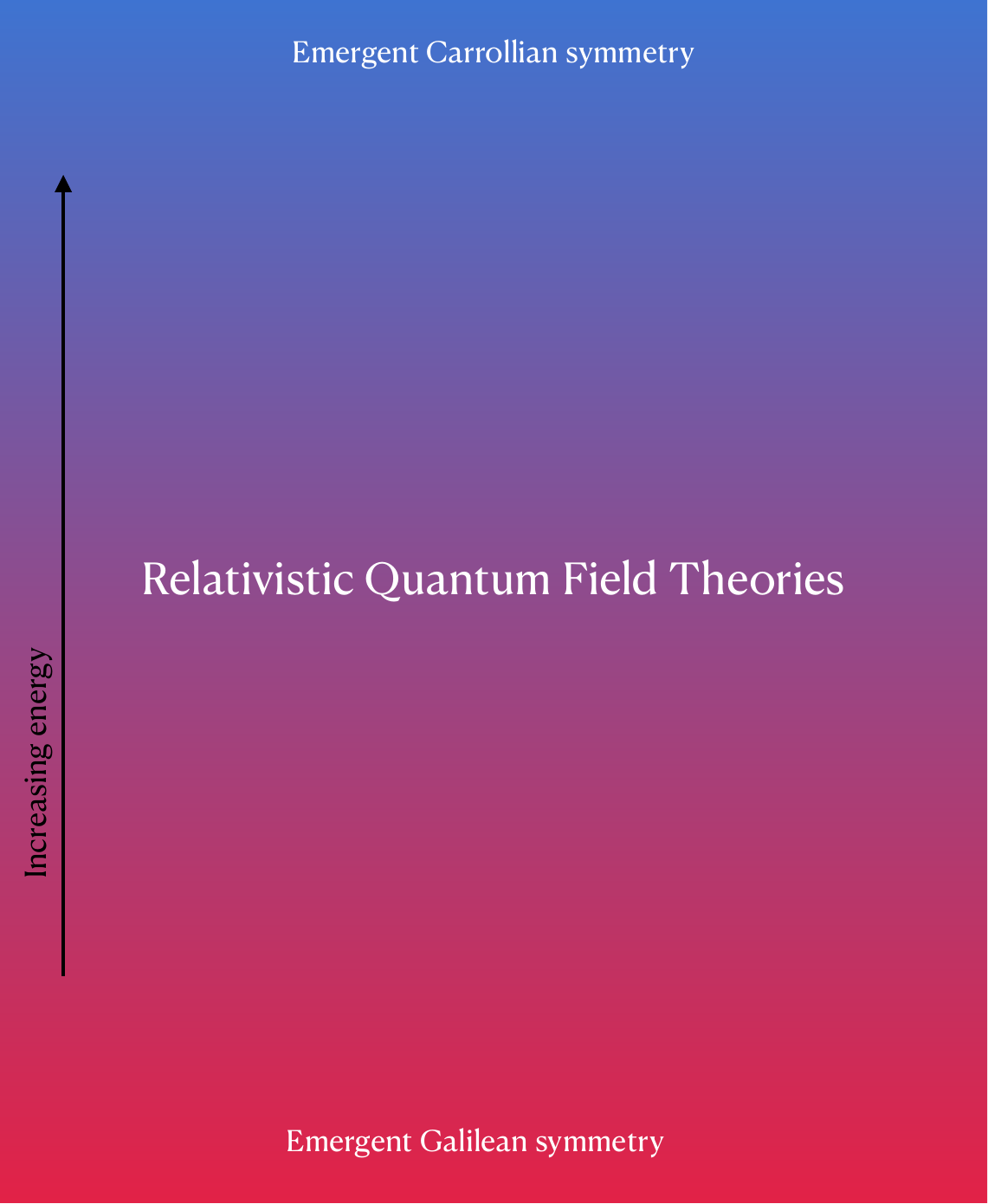}
\caption{Emergent symmetries in relativistic QFTs}
\label{fig:UVIR}
\end{figure}
As mentioned above, moving up the RG flow is fraught with danger, and irrelevant operators, which would trigger such flows, also necessitate the addition of more and more counter terms to the Lagrangian of the QFT in question to render it finite. A loss of locality is a natural and problematic consequence. Now let us think of the very high energy theory instead of the very low energy one where we argued that Galilean invariance should naturally appear. The Galilean algebra in $d$ dimensions arises as an In{\"o}n{\"u}-Wigner contraction of the Poincare algebra when the speed of light $c \to \infty$, and the non-zero commutators read:
\bea\label{gali}
[J_{ij} , B_k ] &=\delta_{k[j}B_{i]},~~~
[J_{ij},\, P_k] = \delta_{k[j}P_{i]},~~ [H, B_i] = P_i.
\ena
where $[ J_{ij}, J_{kl} ] $ gives the usual spatial rotation algebra in one less dimension, $P_i$ are spatial momenta and $B_i$ are the boosts, with $H$ being the Hamiltonian. As a major point of difference from the Poincare algebra, notice here that Galilean boosts commute
\be
[B_i,B_j] = 0
\ee
as opposed to Lorentzian boosts which close to rotations $[ J_{0i}, J_{0j} ]=J_{ij}$. 

\medskip 

There exists a diametrically opposite limit of the Poincare algebra where the speed of light, instead of going to infinity, goes to zero. This apparently non-sensical $c\to0$ limit of the Poincare algebra has been named the Carroll algebra \cite{LBLL, NDS}, after Lewis Carroll, the creator of Alice in Wonderland \cite{3179}, where the commutators of interest (except spatial rotations, which obey the same algebra as in the Lorentz or Galilean case) are:
\be\label{carrollalg}
[J_{ij}, B_k ]=\delta_{k[j}B_{i]}, ~ [J_{ij}, P_k ]=\delta_{k[j}P_{i]}, ~ [B_i,P_j]=-\delta_{ij}H.
\ee
Carroll boosts again commute. But we see a reversal of roles between the spatial momenta and the Hamiltonian in this algebra compared to the Galilean algebra above. We claim that the very high energy sector of a relativistic quantum field theory is governed by the Carroll algebra, in a manner similar to the expectation that the Galilean algebra dictates the very low energy sector. We now justify our rather radical claim. 

\medskip

The very high energy theory may be thought of, in the spirit of Wilson, as a theory where all the low energy modes are inconsequential. It is conceivable that such a theory may be reached from a lower energy theory by systematically throwing away or ``integrating out" all modes except the very high energy ones. In a sense we are envisioning this extreme high energy theory as being built out of a sector of the original theory, very much like the very low energy theory. This very high energy theory should be limited to excitations that travel extremely fast and as a first approximation, can be thought of only containing modes which travel at the speed of light. This in a sense means that the theory should contain only excitations moving on null hypersurfaces in the spacetime on which the theory was originally defined. Now, it is known that Carroll symmetries appear on generic null hypersurfaces and hence it can be expected that the very high energy sector is governed by Carrollian symmetries. 

\medskip

We now offer more evidence that shows our claim may not be as radical as may have been assumed at first. We will do this by showing the emergence of Carrollian symmetries in various ultra-high energy contexts. But before we go into specific examples, let us offer some general explanations. The $c\to0$ limit can be viewed as a limit where 
\be\label{limit1}
t \to \varepsilon t, \, x^i \to x^i, \quad \varepsilon\to 0,
\ee
where $\varepsilon$ is a dimensionless parameter which plays the role of the speed of light. Now the contraction of the time direction essentially means that the energies become much larger than the spatial momenta, i.e.
\be{}
E \to E/\varepsilon, \, p^i \to p^i
\ee
So, the Carroll limit naturally takes the theory to high energies. Below we exemplify this in various contexts and show how Carrollian structures arise in the ultra-high energy sectors of various theories. 

\subsection{Carroll and extreme high energies: Examples}
We now give specific examples of how Carrollian symmetries arise in the extreme high energy limits of various theories of interest. 
\begin{itemize}
\item[$\square$] {\em Tensionless strings:} Non-interacting string theory is most conveniently described in terms of the Polyakov action:
\be{}
S_P = \frac{1}{2 \pi \alpha'} \int d^2\xi \sqrt{\gamma} \gamma^{ab} \partial_a X^\mu \partial_b X^\nu \eta_{\mu\nu}
\ee
In the above, $\gamma^{ab}$ is the metric on the worldsheet, while $ \eta_{\mu\nu}$ is the flat background metric indicating that the string moves in a flat spacetime geometry. The theory has one tuneable parameter $\alpha'$, the length of the fundamental string. The Polyakov action has diffeomorphism and Weyl invariance which need gauge fixing. In the so-called conformal gauge, where the worldsheet metric is fixed to be flat, there is still a residual gauge symmetry (diffeomorphisms that can be undone by a Weyl transformation) which leads to conformal invariance of the worldsheet theory. This helps organise the spectrum and indeed the quantum theory of the tensile relativistic string. 

String theory reduces to Einstein gravity (supersymmetric version thereof) when the length of the string goes to zero, i.e. $\alpha' \to 0$. This is the low-energy limit of string theory. The opposite limit, the ultra-high energy limit is where $\alpha' \to \infty$. Since the tension is the inverse of the string length, this is also the tensionless limit, where the string becomes long and floppy. In this limit the worldsheet metric degenerates and the action can be written as \cite{Isberg:1993av}
\be{}
S_{\text{ILST}} = \int d^2\xi \, V^a V^b \partial_a X^\mu \partial_b X^\nu \eta_{\mu\nu}
\ee
where $V^a$'s are vector densities and $V^a V^b$ replaces $\sqrt{\gamma} \gamma^{ab}$ as the tension vanishes. Carrollian structures thus appear on the worldsheet. The residual symmetries in the analogue of the conformal gauge change from the two copies of the Virasoro to the 2d conformal Carroll algebra  \cite{Bagchi:2013bga,Bagchi:2015nca, Bagchi:2020fpr} indicating this change of worldsheet geometry. 

The very high energy regime of string theory thus indicates the emergence of Carrollian structures, in this case on the worldsheet of the string, which becomes tensionless in this sector. 

\item[$\square$] {\em Ultra-relativistic fluids:} The soup called Quark-Gluon-Plasma (QGP) formed immediately after the Big Bang is effectively modelled by fluids flowing at velocities near the speed of light. One of the simplest and most useful models in this scenario was put forward by Bjorken \cite{Bjorken:1982qr} and this effectively two dimensional model goes under the name of Bjorken flow. It has been recently shown that this model admits Carrollian symmetry \cite{bjorken1}. A generalisation of this model to include dependences in the transverse directions is the so-called Gubser flow \cite{Gubser:2010ze}. It was shown even more recently that Carroll hydrodynamics can effectively describe even Gubser flows \cite{bjorken2}. In general, Carrollian hydrodynamics should be able to capture all ultra-relativistic fluids flowing at velocities comparable to the speed of light. The very high energy sector of fluid dynamics hence is effectively captured by a Carroll sector. 

\item[$\square$] {\em Flat space holography:} Holography beyond AdS/CFT and in particular flat space holography has got a lot of attention of late and progress has been achieved in two slightly differing approaches called Celestial holography (see \cite{Strominger:2017zoo, Pasterski:2021rjz, Raclariu:2021zjz} for excellent reviews) and Carrollian holography. See \cite{Bagchi:2010zz,Bagchi:2012cy,Bagchi:2012xr,Barnich:2012xq,Bagchi:2015wna, Bagchi:2014iea,Bagchi:2016bcd} for earlier work in the 3d bulk/2d boundary context. More recent developments for 4d bulk and 3d boundary have followed from \cite{Bagchi:2022emh,Donnay:2022aba}. The main difference between the two is that while the Celestial version advocates a co-dimension two hologram in terms of a relativistic conformal field theory living on the celestial sphere on the null boundary of flat space, the Carroll formulation keeps track of the null direction at $\mathscr{I}^\pm$ and hence puts forward a co-dimension one dual. As the name suggests, Carrollian holography posits that 4d asymptotically flat spacetime has a holographic dual which is a 3d Carrollian CFT. See \cite{Bagchi:2023cen} for a discussion on the points of difference between the approaches. 

Now, flat spacetimes can be obtained by an infinite radius limit of AdS. This limit can also be thought of as focusing on length scales very small compared to the radius of AdS deep inside the bulk where the effect of curvature would not be felt. In terms of the boundary theory, this region corresponds to the deep UV region of the CFT. Hence flat holography can be thought of as a very high energy sector of usual AdS/CFT. Now, it has been recently made precise that this limit on AdS/CFT gives rise to the Carrollian hologram described above \cite{Bagchi:2023fbj,Bagchi:2023cen}. This was done by showing how various AdS Witten diagrams in the limit led to Carrollian correlation functions. Again, the emergence of Carrollian structures, now in the very high energy sector of AdS/CFT is indicative of the fact that Carrollian symmetry may in general be associated to the deep UV of a generic quantum field theory and this symmetry may be emergent in the same way as Galilean symmetries emerge in the very low energy sector of any relativistic set up.    

\end{itemize}

Having offered several examples and hopefully having convinced the reader that the initial claim of the emergence of Carrollian symmetry in the ultra-high energy sector of any relativistic theory is not unnatural, we make a final comment before outlining how we will go about showing this in the context of RG flows in QFTs in this paper. The operators we naturally associate with flows from the IR to the UV are irrelevant operators and we have remarked earlier that perturbation of a QFT by generic irrelevant operators would lead to a loss of locality in the theory. It seems from our discussion above that somewhere in the extreme UV, the relativistic theory morphs into a Carrollian theory. One would expect locality (in the sense that it is defined in the relativistic regime) to be destroyed in this Carrollian theory since the most ``natural" way of reaching this regime would be through irrelevant deformations. The loss of locality (defined in the relativistic sense) is manifest in the Carroll theory since lightcones completely close up in this regime. If anything, the theory becomes ultra-local.  

\subsection{What is in this paper?}

Having set the stage in the previous subsections, our aim in this paper is to show how Carrollian symmetry indeed appears in quantum field theories through renormalisation group flows. While an important future goal is to show that this happens generically, in the present work we will be very specific and will focus on two dimensional relativistic conformal field theories and perturbation of 2d CFTs by marginal operators. We will show that when we dial some effective parameter controlling the size of the perturbation all the way to infinity so as to reach the edge of the parameter space, we indeed find the emergence of Carrollian structures. In this case, the marginal deformations which generate continuous lines of flows from one to another CFT end up in a 2d Carrollian CFT at the end of this process. 

\paragraph{Outline of the paper:} Sec.\eqref{sec2} contains a summary of our results and this should be useful for readers who wish to just get the gist of the paper without ploughing through the technicalities. In particular, Table \eqref{tab1} contains the most vital information which we would go on to show in detail in the later sections. The two following sections contains the main details of the paper. Sec.\eqref{sec3} deals with the symmetric current-current deformations. We first address the classical aspects and then move into the quantum mechanical aspects later in the section. Sec.\eqref{sec4} addresses the antisymmetric deformation and again we delve into both classical and quantum mechanical details of the deformed theory. The two deformations are shown to act very differently, but as advertised above, they both lead to theories with Carrollian conformal symmetry at the end of the flow. We end with comments and future directions in Sec.\eqref{sec5}. A number of appendices at the end contain calculations omitted in the main text and supplementary discussions.  

\newpage

\section{Flowing to Carroll: the set up}\label{sec2}
\subsection{Summary of results}
In this brief section, we will outline our strategy for the rest of the paper. The theory we will concentrate on is the free massless relativistic scalar theory in $d=2$. This is of course a relativistic conformal field theory. We will, in general, consider $D$ such scalar fields:  
\be
\cL=\frac{1}{2} \sum_{i=1}^{D}\left(\left(\partial_\tau \phi_i\right)^2-\left(\partial_\sigma \phi_i\right)^2\right).
\ee
which gives rise to the equations of motion for each field:
\be
	\left(\pd_\tau^2-\pd_\sigma^2\right)\phi_i=0.
\ee
We will perturb this theory in two different ways, both of them will involve adding bilinears of $U(1)$ Kac-Moody currents.
\begin{itemize}
\item[$\star$] {\em Symmetric current-current deformation:} We first look at deformations of the parent Lagrangian by a symmetric current-current deformation:
\be \label{lag.deform1}
\widetilde{\cL}^\a=\cL+\frac{\a}{2}\sum_i \left({{J}}^2_{i}+\overline{{J}}^2_{i}\right).
\ee
Here $J,  \overline{{J}}$ are conserved currents: 
\be
	{J}_i=\frac{1}{\sqrt{2}}(\dot{\phi}_{i}+ \phi_{ i}^{\prime}),\quad \overline{{J}}_{i}=\frac{1}{\sqrt{2}}(\dot{\phi}_{i}- \phi_{i}^{\prime}),
\label{JbarJ.currents}
\ee
where $\dot{\phi}_i=\pd_\tau{\phi}_i$ and ${\phi}'_i=\pd_\sigma{\phi}_i$ are the temporal and spatial derivatives, and $\a$ is a continuous parameter.
The currents obey the conservation laws $\bar{\pd}J_i=\pd \overline{J}_i=0$ with $\pd, \bar{\pd}=\pd_\tau\pm\pd_\sigma$. More explicitly, the deformed Lagrangian reads:
\be\label{JJdef}
	\widetilde{\cL}^\a=\frac{1}{2} 
	\sum_{i=1}^{D}\left((1+\a)\left(\partial_\tau \phi_i\right)^2-(1-\a)\left(\partial_\sigma \phi_i\right)^2\right).
\ee
The deformation scales the time derivatives up in the action and scales down space derivatives, or vice versa, depending on the chosen sign of $\a$. When $\a\neq \pm 1$, one can rescale temporal and spatial coordinates and ``restore'' conformal symmetry. 
$\a = \pm 1$ acts as boundaries of the parameter space,  at  which one of the derivatives in the action is lost. We will show that $\a=1$ is the Carroll point (and $\a=-1$ the Galilean point) and hitting this leads to a change of space-time symmetry. Both from the classical and quantum perspective, this is the exact limit where Carroll algebra takes over.
\medskip

The unequal scaling of spacetime coordinates as manifest in action \eqref{JJdef} leads to a modification of the effective speed of light in the theory. Using this handle over the propagation velocity of modes, we will tune the effective speed of light to zero, where only fast-moving excitations are important. In \cite{Bagchi:2022nvj}, similar conclusions were reached via a ``boost'' transformation on the (anti)holomorphic currents, parameterized by $\a$. We will show that this effectively leads to a Bogoliubov transformation on the modes of the theory, and different values of $\a$ define different physical vacua on which a line of CFTs are defined, albeit with more and more faster moving excitations. 


\begin{table}[t]
\begin{tabular} {|c|c|c|}
\hline
Property & Symmetric & Antisymmetric \\
\hline
Deformed Lagrangian & $\widetilde{\cL}^\a=\cL+\frac{\a}{2} \left({{J}}^2_{i}+\overline{{J}}^2_{i}\right)$ & $\widetilde{\cL}^g=\cL+ g \epsilon_{ab}J^a\bar J^{b}$ \\
Deformed Hamiltonian & $\widetilde{\cH}^\a=\cH \left[\pi(\a),\phi\right]+\a J_i \overline{J}_i $ & $\widetilde{\cH}^g=\cH\left[\pi(g),\phi\right]+ g \epsilon_{ab}J^a\bar J^{b}$ \\
End-point of flow & $\a \to \pm 1$ & $g\to \infty$ \\
Symmetry at end-point & BMS$_3$ & BMS$_3$ \\
Speed of light & Changes continuously to $0$ & Abruptly $\to 0$ @ end-point \\
Vacuum & Continuous change& Sudden jump @ end-point \\
Central charges & Continuous change & Sudden jump @ end-point \\
Nature of flow & Bogoliubov transformation driven & Spectral flow driven\\
\hline
\end{tabular}
\caption{Properties of deformations.}
\label{tab1}
\end{table}

\item[$\star$] {\em Antisymmetric current-current deformation:}
We will then focus on current-current deformations of the form $\epsilon_{ab}J^a\bar J^{b}$. These are exactly marginal operators, and have been discussed in the literature in connection to string theory \cite{Apolo:2019zai}. A salient feature of such deformations is that these mix spacetime indices on the fields, as opposed to the symmetric $J\bar{J}$ case where spacetime indices just decouple from each other. 
\medskip

We will restrict ourselves to the two dimensional Lagrangian density of two bosonic fields $\phi_{1,2}$ deformed by a $J^1\wedge \bar{J}^2$ operator for simplicity. Generalisation to $2D$ bosons is immediate. Here again the currents $J^i = \partial_z\phi^i$ and $\overline{J}^i = \partial_{\bar{z}}\phi^i$ are the (anti-)holomorphic $U(1)$ currents of the theory. 
The deformed Lagrangian, as discussed in \cite{Guica:2021fkv}, is
\be\label{lagwedge}
	\mathcal{\tilde{L}}^{(g)}=\frac{1}{1+g^2}\left[\frac{1}{2} 
	\sum_{i=1,2}\left(\left(\partial_\tau \phi_i\right)^2-\left(\partial_\sigma \phi_i\right)^2\right)
	-g\left(\partial_\tau \phi_1 \partial_\sigma \phi_2-\partial_\sigma \phi_1 \partial_\tau \phi_2\right)\right].
\ee
The action of this deformation is very different from the symmetric case. For all finite values of the coupling $g$, the theory is a relativistic CFT, as the perturbation term can be considered a boundary contribution. This $g$-deformed line of CFTs however continue to have the same effective speeds of light, as the equation of motion for propagating modes remain the same. The change of spacetime algebra only becomes apparent when this boundary term becomes unmistakably large at $g\to \infty$. We will show how a non-invertible change in boundary conditions for the fields in this limit implies zero velocity of mode propagation, thereby bringing about an abrupt change in spacetime algebra. 
\medskip

Quantization of the theory needs careful consideration of the boundary term contribution. There is no change in physical vacuum of the theory at finite $g$. However the gradual take-over by higher and higher energy modes is achieved by a process called the ``spectral flow" of energy eigenstates. The bare conformal dimensions of undeformed fields pick up $g$ dependent contributions as the strength of the perturbation increases. 

For finite $g$, these deformed states continue to transform under Virasoso-Kac-Moody symmetries, as the (momentum dependent) spectral flow parameter commutes with all the generators. Starting with the highest weight representations of the undeformed Virasoro algebra, the CFT Hilbert space reorganizes itself at the strict $g\to \infty$ limit into the spectrum of the Carrollian conformal algebra. 
\end{itemize}

From our brief summary above, it should be clear to the reader that although the two deformations take us from the same relativistic theory to a theory with Carrollian symmetries, the energy flows upwards from the relativistic regime as triggered by different operators have very different effects on the theories, as elucidated in Table \eqref{tab1} as well. This is the key idea that we will delve into in our work.

\subsection{Carrollian limits and the Dirac-Schwinger condition}

Before we begin our investigations of the edges of parameter spaces of QFTs, we first need to acquaint ourselves with some properties of Carrollian theories and the limits which lead up to these theories. As discussed in the introduction, the Carroll algebra arises out of a $c\to0$ limit of the Poincare algebra and is given by \eqref{carrollalg}. We are interested in quantum field theories exhibiting Carrollian (and Conformal Carrollian) symmetries (see \cite{Bagchi:2016bcd,Bagchi:2019clu,Bagchi:2019xfx,Hao:2021urq, Basu:2018dub,Saha:2022gjw, Banerjee:2023jpi,deBoer:2023fnj} for a non-exhaustive list). Following the algebra, an obvious way to get to Carrollian QFTs is to dial $c\to0$ in particular relativistic QFTs. Below we switch to the Hamiltonian framework to elucidate some properties of this limit. 

\medskip

To understand how Carrollian dynamics occurs in field theories, we now revisit Carrollian limits from the point of view of the Dirac-Schwinger condition of classical field theory. We will be mostly following the notations introduced in \cite{Henneaux:2021yzg} in what follows.  Let us take the example of a massless relativistic scalar field on flat D-dimensional Minkowski space, and reinstate the $c$ factors so that we can clearly analyze the effect of taking $c\to0$ limit. Here the Lagrangian is given by:
\be\label{lagra}
\cL = \frac{1}{2c^2}\left(\dot\phi^2 - c^2 (\nabla_k \phi)^2\right).
\ee
As discussed before, Carrollian limit corresponds to taking $c\to 0$ limit on the theory.  So naively, the limiting theory would be the one where the temporal derivative term leads over the spatial derivative, and the resulting theory becomes invariant under Carroll boosts. However, one could also have a subtle alternative limit, where spatial gradients survive\footnote{Note that in the opposite limit $c\to \infty$, just the spatial derivative part of the action survives, and becomes invariant under Galilean boosts instead, as discussed earlier. This is subtly different from this alternative Carroll limit with spatial gradients.}. These sectors will also be evident in the current work, albeit we will mostly use the convention $c=1$ and the Carroll limit will be imposed via dialing of some auxiliary parameters as promised in the previous subsection.  A nice way to understand these two Carroll limits is to consider the Hamiltonian formalism, where energy (Hamiltonian) density $\cH$ and momentum density $\cP_k$ form the local generators of the Carroll symmetry algebra. The Hamiltonian density corresponding to the scalar field is:
\be
\cH = \frac{1}{2}\left(c^2\pi^2+ (\nabla_k \phi)^2\right),~~\dot{\phi} = \frac{\pd \cH}{\pd\pi}
\ee
and the invariant action reads
\be
S = \int dt~d^dx \left(\pi\dot\phi-\cH\right).
\ee
For a relativistic scalar field, the Poisson bracket of energy densities at different spatial points generate the Dirac-Schwinger condition:
\be
\left\{\mathcal{H}(x), \mathcal{H}\left(x^{\prime}\right)\right\}_{\mathrm{PB}}
=2 \cP^k(x) \partial_{k}\delta\left(x-x^{\prime}\right)
+\left(\partial_k \cP^k(x')\right) \delta\left(x-x^{\prime}\right),
\ee
where one uses the canonical commutation relations
\be
	\{\phi(x),\pi(x')\}_{\rm PB}
	=-\{\pi(x),\phi(x')\}_{\rm PB}
	=\delta(x-x'),
\ee
and the momentum flux density is given by $\cP_k = c^2\pi \nabla_k\phi$. 

\medskip

\noindent The hallmark of Carroll dynamics is the vanishing of the Poisson bracket between Hamiltonian densities, since the Carrollian momentum flux density is identically zero. So for us, the rule of thumb to construct Carroll invariant systems would be to look for the ones having
\be\label{HPB}
\{\mathcal{H}(x),\mathcal{H}(x')\}_{\rm PB} = 0.
\ee
Consequently one can generalise this statement to the quantum case where the commutator vanishes. 
\footnote{This condition can also be dubbed as the ``flat band'' condition as used by condensed matter physicists, and a striking connection to Carroll physics to such systems was shown in \cite{Bagchi:2022eui}. }
The simplest example of such a system corresponds to taking $c\to 0$ directly in \eqref{lagra}, as mentioned beforehand, where only the time derivative part of the action remains. This version of Carrollian limit was dubbed as the ``Magnetic'' limit since $\mathcal{H}$ just contains the spatial gradient, and the momenta cannot be expressed in terms of the derivative of fields. On the other hand one could canonically rescale the fields and momenta using $\phi \to c\phi$ and $\pi \to \frac{1}{c}\pi$, and then take $c\to 0$, leading to 
 $\mathcal{H}\sim \pi^2$. This other limit is called the ``Electric" Carroll limit, and in this case equations of motion simplify to $\pi = \dot\phi$ and $\dot\pi = 0$. The nomenclature is sort of opposite in spirit to similar Carroll limits taken from the Lagrangian perspective that we started out with. See \cite{deBoer:2021jej} for more details on Carroll expansions of scalar fields and related boost invariance . 
 \medskip
 
In both kind of limits \eqref{HPB} however is sacrosanct. This brings forth the possibility that for a multi-field system one might have a myriad of ways of taking Carroll limits, and even a mixture of electric and magnetic Carroll fields may be admissible. This could lead to very complicated phase space structures, reducing it down to a combination of a subset over volume forms. In the rest of this manuscript, we will be talking about deformation of scalar field theories via current bilinears, that exactly achieve different kinds of Carroll limits on the system as their coupling saturates some relevant bound. Upon quantization this would mean the resulting spectrum approaches the Carrollian one via different paths in the space of couplings. We would be consistently switching between Lagrangian and Hamiltonian languages to get these points through to the reader. 

\bigskip

\section{The symmetric deformation }\label{sec3}

We begin our detailed study of deformations with the symmetric current-current deformation of the scalar theory:  
\be\label{defL}
\widetilde{\cL}^\a=\cL_\text{free} + \mathcal{O}_\text{def} = \frac{1}{2} \sum_{i=1}^{D}\left(\left(\partial_\tau \phi_i\right)^2-\left(\partial_\sigma \phi_i\right)^2\right) +\frac{\a}{2}\sum_i \left({{J}}^2_{i}+\overline{{J}}^2_{i}\right).
\ee
where as stated before ${J}_i=\frac{1}{\sqrt{2}}(\dot{\phi}_{i}+ \phi_{ i}^{\prime}),\, \overline{{J}}_{i}=\frac{1}{\sqrt{2}}(\dot{\phi}_{i}- \phi_{i}^{\prime})$ are conserved currents. 

\subsection{Classical considerations}
We will find it convenient to move to the Hamiltonian formalism, especially for the purposes of quantizing the theory. We first set up some notation. 
\subsubsection{Setting the stage}
The canonical commutation relation in this case reads
\be
	\{\phi_i(\sigma),\pi_j(\sigma')\}_{\rm PB}
	=-\{\pi_i(\sigma),\phi_j(\sigma')\}_{\rm PB}
	=\delta_{ij}\delta(\sigma-\sigma'),
\ee
with all other brackets vanishing. Here $\pi_i$ are the conjugate momenta for the fields $\phi_i$, given by
\be
	\pi_i=\frac{\pd \cL_\text{free}}{\pd \dot{\phi}_i}=\dot{\phi}_i.
\ee
These relations imply the algebra among the currents
\begin{align}
	\{J_i(\sigma),J_j(\sigma')\}_{\rm PB}
	=\delta_{ij}\pd_{\sigma}\delta(\sigma-\sigma'), 
	\quad 
	\{\overline{J}_i(\sigma), \overline{J}_j(\sigma')\}_{\rm PB}
	=-\delta_{ij}\pd_{\sigma}\delta(\sigma-\sigma'),
\end{align}
with $J_i$ and $\overline{J}_j$ Poisson commuting. The (anti)holomorphic stress tensors for the whole system are (classically) obtained via Sugawara construction from the currents
\be
	T=\frac{1}{2}\sum_i {J}_i^2,
	\qquad\quad
	\overline{T}=\frac{1}{2}\sum_i \overline{J}_i^2,
\label{JbarJ.stress}
\ee
which satisfy the following algebra
\begin{subequations}
\begin{align}
\left\{T(\sigma), T\left(\sigma^{\prime}\right)\right\}_{\mathrm{PB}} 
& =2 T(\sigma) \partial_\sigma \delta\left(\sigma-\sigma^{\prime}\right)
+\left(\partial_\sigma T(\sigma)\right) \delta\left(\sigma-\sigma^{\prime}\right),
\\
\left\{ \overline{T}(\sigma),  \overline{T}\left(\sigma^{\prime}\right)\right\}_{\mathrm{PB}} 
& =-2  \overline{T}(\sigma) \partial_\sigma \delta\left(\sigma-\sigma^{\prime}\right)
-\left(\partial_\sigma  \overline{T}(\sigma)\right) \delta\left(\sigma-\sigma^{\prime}\right),
\\
\left\{T(\sigma), \overline{T}\left(\sigma^{\prime}\right)\right\}_{\mathrm{PB}} 
& =0.
\end{align}
\end{subequations}
As a consequence, when we expand the stress tensors in terms of Fourier modes:
\be\label{Lmodes}
	T(\sigma)=\sum_m \cL_m\,e^{-im\sigma}
	\qquad\quad
	\overline{T}(\sigma)=\sum_m \overline{\cL}_m\,e^{+im\sigma},
\ee
the generators $\cL$'s correspond to two copies of (the classical part of) the Virasoro algebra. 
The above Poisson brackets between stress tensors can then be rewritten in terms of these Fourier modes:
\be\label{Vir1}
\{\cL_n, \cL_m\}_{\text{PB}} = -i(n-m) \cL_{n+m}, \qquad \{\bL_n, \bL_m\}_{\text{PB}} = -i(n-m)\bL_{n+m},
\ee
which are the well known form of centerless Virasoro or Witt algebra. In the quantum case, one will also have central charges $c,\bar{c}$ appearing in the commutators of these generators.
\subsubsection{BMS at the end of parameter space}
With our notations in place, we now show how the spacetime algebra would change at the ends of the flow generated by this deformation. 
Let us now focus on the conjugate momenta to the fields as calculated from \eqref{defL}:
\be
	\widetilde{\pi}^\a_i=\frac{\pd \widetilde{\cL}^\a}{\pd \dot{\phi}_i}=(1+\a)\dot{\phi}_i
\label{JbarJ.conjm}
\ee
Thus the Hamiltonian density can be written as
\be\label{JbarJ.ham}
	\widetilde{\cH}^\a
	=\sum_{i=1}^{D}\widetilde{\pi}_i\dot{\phi}_i-\widetilde{\cL}^\a
	=\frac{\k}{2}
	\sum_{i=1}^{D}\left(\e\,\widetilde{\pi}_i^2+\frac{1}{\e}\left(\partial_\sigma \phi_i\right)^2\right)
\ee
where we have defined new parameters
\be\label{kappa}
	\k=\sqrt{\frac{1-\a}{1+\a}}, 
	\qquad
	\e=\frac{1}{\sqrt{1-\a^2}}.
\ee
These would be of interest in what follows. Now, the deformed Hamiltonian \eqref{JbarJ.ham} written in terms of the undeformed one is
\be\label{hami}
\widetilde{\cH}^\a= \cH_\text{free} +2\a \sum_i  {{J}}_{i} \overline{{J}}_{i}=T + \overline{T}+2\a \sum_i  {{J}}_{i} \overline{{J}}_{i}.
\ee
where $\cH_\text{free}$ is just the Hamiltonian of $D$ free massless relativistic scalars and $T$ and $\overline{T}$ are stress tensors defined above. Along with this, we also define 
\be
\cJ=T - \overline{T}.
\ee
Following \cite{Bagchi:2022nvj}, we now calculate the algebra of $\widetilde{\cH}^\a$ and $\cJ$ using the current algebra as mentioned in the last section. This leads us to the following algebra:
\be\label{flowalgebra}
\begin{aligned}
& \left\{\mathcal{J}(\sigma), \mathcal{J}\left(\sigma^{\prime}\right)\right\}_{\mathrm{PB}}
=2 \cJ(\sigma) \partial_\sigma \delta\left(\sigma-\sigma^{\prime}\right) 
+\left(\partial_\sigma \cJ(\sigma)\right) \delta\left(\sigma-\sigma^{\prime}\right)\\
& \left\{\mathcal{J}(\sigma), \widetilde{\cH}^\a\left(\sigma^{\prime}\right)\right\}_{\mathrm{PB}}
=2 \widetilde{\cH}^\a(\sigma) \partial_\sigma \delta\left(\sigma-\sigma^{\prime}\right)
+\left(\partial_\sigma \widetilde{\cH}^\a(\sigma)\right) \delta\left(\sigma-\sigma^{\prime}\right) \\
& \left\{\widetilde{\cH}^\a(\sigma), \widetilde{\cH}^\a\left(\sigma^{\prime}\right)\right\}_{\mathrm{PB}}
=\left(1-{\alpha}^2\right)
\left(2 \cJ(\sigma) \partial_\sigma \delta\left(\sigma-\sigma^{\prime}\right)
+\left(\partial_\sigma \cJ(\sigma)\right) \delta\left(\sigma-\sigma^{\prime}\right)\right).
\end{aligned}
\ee
Written in terms of modes
\be\label{LMmodes}
	\cJ(\sigma)=\sum_m L_m\,e^{-im\sigma},
	\qquad
	\widetilde{\cH}^\a(\sigma)=\sum_m M^\a_m \, e^{-im\sigma},
\ee
we have the algebra
\begin{align}
\{L_n, L_m\}_{\text{PB}} &= -i(n-m) L_{n+m}, \quad \{L_n, M^\a_m\}_{\text{PB}} = -i(n-m)M^\a_{n+m},  \cr
\{M^\a_n, M^\a_m\}_{\text{PB}} &= \left(1-{\alpha}^2\right)\Big[-i(n-m) L_{n+m}\Big].
\end{align}
Here $\a=0$ is the usual conformal algebra in two dimensions. For $-1<\a<1$, we can always rescale the $\widetilde{\cH}^\a$ by a constant factor and redefine the above as the usual conformal algebra. 

\medskip

The interesting thing happens at the edge of the parameter space, given by $\a=\pm1$. The above procedure of rescaling fails at these end points and the last bracket $\{\cH, \cH\}$ vanishes. The spacetime algebra deforms from the two copies of the Virasoro algebra to  (the classical part of) the BMS$_3$ algebra  \cite{Bondi:1,Sachs:1962zza}\footnote{The last condition is nothing but smooth vanishing of the Dirac-Schwinger condition, as described earlier. This can also be interpreted as a $SL(2,\mathbb{R})$ transformation on the Kac-Moody currents that degenerates in certain limits.}. The BMS$_3$ algebra is isomorphic to the Conformal Carrollian Algebra in $d=2$ (CCA$_2$) and hence we see the emergence of Carrollian symmetry at the very end of parameter space. We will see below that $\a=1$ is the Carroll point where the effective speed of light goes to zero.
\footnote{It is imperative to point out here that the generators of the BMS$_3$ algebra ($L,M$) can be constructed by an In\"on\"u-Wigner contraction of two copies of undeformed virasoro generators as in \eqref{Vir1}:
\be\label{IW}
L_n = \cL_n - \overline{\cL}_{-n},~~~M_n = \varepsilon\left( \cL_n + \overline{\cL}_{-n}\right),~~\varepsilon\to 0,
\ee
which is the consequence of the ultrarelativistic limit \eqref{limit1}. Despite using same notations, the above contraction of $M$ is certainly different than what \eqref{hami} implies at the level of modes. 
}  
\medskip

There is also the other end $\a=-1$ where the speed goes to infinity and this is the Galilean point. The emergence of the same algebra both deep in the UV and deep in the IR of the original relativistic CFT points to a curious feature of $d=2$ where the Carrollian and Galilean algebras, as well as their conformal extensions are isomorphic \cite{Bagchi:2010zz}. We will deal mainly with the Carrollian very high energy theory in this paper and leave the details of the Galilean low energy theory for future work. 

\paragraph{Dialling the speed of light.} We now wish to identify the parameters we introduced to the speed of light in the system and thereby connect to Carrollian symmetries in a more intuitive fashion. Since the canonical relations are set by
\be
\label{can.rel}
	\{\widetilde{\pi}_i(\sigma),\phi'_j(\sigma')\}_{\rm PB}
	=\{\phi'_i(\sigma),\widetilde{\pi}_j(\sigma')\}_{\rm PB}
	=\delta_{ij}\pd_{\sigma}\delta(\sigma-\sigma')
\ee
and otherwise vanishes,
the time evolution of the system is given by the following Hamiltonian equations of motion:
\bea
	\dot{\phi}_i(\sigma)
	&=\{\phi_i(\sigma),\widetilde{H}\}_{\rm PB}
	\qquad\qquad\left(\widetilde{H}=\int d\sigma'\,\widetilde{\cH}(\sigma')\right)\nn
	&=\int d\sigma'\,
	\sum_{j=1}^{D}\frac{\widetilde{\pi}_j(\sigma')}{1+\a}\{\phi_i(\sigma),\widetilde{\pi}_j(\sigma')\}_{\rm PB}\nn
	&=\int d\sigma'\,
	\sum_{j=1}^{D}\frac{\widetilde{\pi}_j(\sigma')}{1+\a}\delta_{ij}\delta(\sigma-\sigma')
	=\frac{\widetilde{\pi}_i(\sigma)}{1+\a}
\ena
which is nothing but reproducing the relation \eqref{JbarJ.conjm}. In a similar manner, 
\be
	\dot{\widetilde{\pi}}_i(\sigma)
	=\{\widetilde{\pi}_i(\sigma),\widetilde{H}\}_{\rm PB}
	=(1-\a)\phi''_i(\sigma).
\ee
Hence the scalar fields in the deformed theory obey the wave equation with an ``effective speed of light"  $\k$,
\be\label{waveeqndef}
	\ddot{\phi}_i-\k^2{\phi}''_i=0,
\ee
where $\k$ is as given in \eqref{kappa}. Hence the BMS points, where $\a=\pm1$, correspond to speed of wave propagation either going to zero (for $\a=1$) or to infinity (for $\a=-1$). At $\kappa = 0$, the Hamiltonian density in \eqref{JbarJ.ham} Poisson commutes with itself, signifying an emergence of Carrollian symmetries. {\footnote{The reader would argue that this is also true at $\kappa\to\infty$ and since the Hamiltonian Poisson commutes with itself here as well, this point also leads to Carrollian and not Galilean symmetries. The issue is a bit more subtle however. The Carroll and Galilei (conformal) algebras may be isomorphic, but the identification of generators is non-trivial and the temporal and spatial directions are flipped in the identification. The Carroll Hamiltonian is no longer the Galilean Hamiltonian, but identified the Galilean momentum. Hence what Poisson commutes is the Galilean momenta and not the Hamiltonian. See Appendix \eqref{appC} for more details on this issue.}} Note that the relative sizes between the momentum and spatial derivative terms in this case is controlled by the effective parameter $\epsilon$ \eqref{JbarJ.ham}, which runs off to infinity.

\subsection{The quantum perspective}
Having spent some time trying to put in place the classical aspects of the effects of the symmetric current-current deformation on free massless scalar theory, we now turn our attention to the quantum aspects. To begin with, we will revisit the previous analysis of  \cite{Bagchi:2022nvj} of rewriting the deformation in terms of a field rotation of underlying currents. We will then use this insight to make statements about first the deformation of the vacuum of the theory and then move on to study the deformation of the spectrum. 

\subsubsection{Deformation as Bogoliubov transformation}
An important observation in \cite{Bagchi:2022nvj} was that the current-current deformation in question \eqref{hami} can be generated by a $SL(2,\mathbb{R})$ rotation of the $U(1)$ Kac-Moody currents associated to the stress tensor of the conformal fields. Writing in a matrix form, this transformation at the level of currents looks like:
\begin{eqnarray}\label{bogo1}
	\begin{pmatrix}
		J^\mu(\ta)\\
		\bar{J}^\mu(\ta)
	\end{pmatrix} = 
	\begin{pmatrix}
		\cosh\theta(\ta) \, \, & \, \, \sinh\theta(\ta)\\
		\sinh\theta(\ta) \, \, & \, \, \cosh\theta(\ta)
	\end{pmatrix}
	\begin{pmatrix}
		J^\mu(0)\\
		\bar{J}^\mu(0)
	\end{pmatrix}
\end{eqnarray}
Here $\ta$ is a boost parameter in the two-dimensional space spanned by (anti-)holomorphic coordinates $z,\bar{z}$. Using the properties of 2d boosts in this space, yields the simple solution
\be
\cosh\theta(\ta) =\frac{1}{\sqrt{1-\ta^2}} ,~~\sinh\theta(\ta)=\frac{\ta}{\sqrt{1-\ta^2}}.
\ee
With this transformation between deformed and undeformed bases, we can now define the combinations of current bilinears:
\begin{subequations}\label{genbog}
\begin{eqnarray}
	\mathcal{J}(\ta;\sigma) &=& \sum_{\mu} \, \big(J^{\mu}(\ta)J^{\mu}(\ta) - \bar{J}^{\mu}(\ta)\bar{J}^{\mu}(\ta)\big) = T(\sigma) - \overline{T}(\sigma)\\
	\mathcal{H}(\ta;\sigma) &=& \frac{1}{(\cosh^2\theta(\ta) + \sinh^2\theta(\ta))}\sum_{\mu} \, \big(J^{\mu}(\ta)J^{\mu}(\ta) + \bar{J}^{\mu}(\ta)\bar{J}^{\mu}(\ta)\big) \nonumber\\
	&=&T(\sigma)+\overline{T}(\sigma) +2\a \sum_{\mu}J^\mu(0)\bar{J}^\mu(0), 
\end{eqnarray}
\end{subequations}
where in the very last line we have re-defined the parameter as\footnote{Note that this redefinition makes sure that $\ta = \pm 1 \implies \a =\pm 1$, consequently $\tanh\theta = \pm 1$.}
\be\label{omega}
\a = \frac{2\cosh\theta\sinh\theta}{\cosh^2\theta + \sinh^2\theta}.
\ee
This is in sync with our discussions earlier around \eqref{hami}, this $\mathcal{H}$ turns out to be our deformed Hamiltonian as expected. We can see with the new parameterisation with $\a$, the stress tensor combinations remain the explicit CFT ones at $\theta = 0~(\a=0)$ and flow over to those of Carrollian algebra at $\theta \to \infty~(\a \to 1)$. In what follows, we will continue to mark all our deformed objects with an $\a$ label, by dropping the tilde notations, to distinguish them from the undeformed ones. 
\medskip

Now let us define Fourier modes associated to the $U(1)$ currents $J,\bar{J}$ as following:
\be{}
J = \sum_{n}j_n e^{-in\sigma}, ~~\bar{J} = \sum_{n}\bar{j}_n e^{+in\sigma}.
\ee
Note that we are working with level one currents $k = \bar{k}=1$. When we quantize, the algebra of currents lead one to the following commutators for modes,
\be{}
[j_n,j_m] = n\delta_{m+n,0},~~[\bar{j}_n,\bar{j}_m] = n\delta_{m+n,0}.
\ee
Consequently, \eqref{bogo1} can be identified as a Bogoliubov transformation between deformed and undeformed modes:
\begin{eqnarray}\label{bogodef}
j_n(\a) = \cosh\theta~ j_n(0) + \sinh\theta ~\bar{j}_{-n}(0);\quad \bar{j}_n(\a) = \cosh\theta~ \bar{j}_n(0) + \sinh\theta~ j_{-n}(0).
\end{eqnarray}
The above relations keeps the canonical commutation relations invariant. We can then follow the usual Sugawara method to construct stress tensors out of bilinears of modes, and that in turn gives rise to the stress tensor modes from \eqref{genbog}:
\begin{subequations}
\begin{eqnarray}
\mathcal{J}_n(\a;\sigma) &=&\sum_{a,m}  \left(j_{-m}^a j_{m+n}^a -\overline{j}_{-m}^a \overline{j}_{m-n}^a\right), \\
\mathcal{H}_n(\a;\sigma) &=& \sum_m  \left(j_{-m}^a j_{m+n}^a+\overline{j}_{-m}^a\overline{j}_{m-n}^a
	+2\a j_{-m}^a\overline{j}_{-m-n}^a\right), 
\end{eqnarray}
\end{subequations}
where the current modes are the undeformed ones $j_n^a = j_n^a(\a=0)$. The algebra of the modes again explicitly leads to the BMS$_3$ algebra when $\a =  1$, as we have discussed in the classical case. Note that this upped bound on $\a$ means the Bogoliubov coefficients diverge in these limits. However, since these modes are related via continuous Bogoliubov transformations, the spectrum also should smoothly interpolate to the BMS spectrum. \footnote{This structure has been particularly well explored within the recent tensionless string literature, see for example  \cite{Bagchi:2015nca, Bagchi:2019cay, Bagchi:2020fpr, Bagchi:2020ats}. In \cite{Bagchi:2021ban} it was shown that this oscillator structures help one understand these theories as those on infinitely accelerated string worldsheets, most noticeably near a black hole horizon.} 
\medskip


\subsubsection{Deformation of the vacuum}
Let us now concentrate on the deformed quantum theory. The highest weight vacuum of the undeformed field theory, as given by 
\be
j_n(0)|0\rangle_{0}= 0 =\bar{j}_n(0)|0\rangle_{0}
\ee 
has now changed due to the Bogoliubov transformed oscillators, and the deformed vacuum condition reads:
\begin{eqnarray}\label{bogovac}
\left(  j_n(0) + \tanh\theta ~\bar{j}_{-n}(0)\right)|0\rangle_{\a}=0; \quad 
 \left(\bar{j}_n(0) + \tanh\theta~ j_{-n}(0)\right)|0\rangle_{\a} =0.
\end{eqnarray}
This means one can write the deformed vacuum as a two-mode squeezed state on the undeformed vacuum:
\be
|0\rangle_\a= {\mathcal{N}} \prod_{n=1}^{\infty}  \exp\left[ \, \frac{1}{n} \tanh\theta~ j_{-n} \cdot \overline{j}_{-n}\right]  |0\rangle_0
\ee
In the above, ${\mathcal{N}}$ is a suitable normalisation factor. At the critical points $\tanh\theta \to \pm 1$ the above state becomes a boundary state, either Dirichlet or Neumann based on the sign of the parameter. \footnote{This class of smoothly interpolating vacua has been considered in the tensionless string literature \cite{Bagchi:2020fpr}.}
\subsubsection{Deformed central charges}
We go forward and elaborate on the  spectrum of the deformed theory now. 
Note that the Bogoliubov transformation of the oscillators lead to a set of new Virasoro generators where the undeformed holomorphic and anti-holomorphic ones mix together:
\begin{eqnarray}
    &&\mathcal{L}_n(\a) = \cosh^2{\theta}\mathcal{L}_n(0) + \sinh^2{\theta}\bar{\mathcal{L}}_{-n}(0) + \frac{1}{2}\cosh{\theta}\sinh{\theta}\sum_p(j_{n-p}\tilde{j}_{-p} + j_{p}\tilde{j}_{-n+p})\nonumber, \\
    &&\bar{\mathcal{L}}_n(\a) = \cosh^2{\theta}\bar{\mathcal{L}}_n(0) + \sinh^2{\theta}{\mathcal{L}}_{-n}(0) + \frac{1}{2}\cosh{\theta}\sinh{\theta}\sum_p(\tilde{j}_{n-p}{j}_{-p} + \tilde{j}_{p}{j}_{-n+p}),\nonumber
\end{eqnarray}
where we have used the standard form of bilinears of oscillators coming from the Sugawara construction and performed a basis transformation. 
Clearly, the interpolating symmetry generator modes of the stress tensors as defined in \eqref{genbog} can be written as combination of these deformed generators,
\begin{eqnarray}\label{GHdef}
\mathcal{J}_n(\a) = \mathcal{L}_n(\a) - \bar{\mathcal{L}}_{-n}(\a), \quad \mathcal{H}_n(\a) = \frac{1}{(\cosh^2\theta + \sinh^2\theta)}\bigg(\mathcal{L}_n(\a) + \bar{\mathcal{L}}_{-n}(\a)\bigg),
\end{eqnarray}
so that we have $\mathcal{J}(\a;\sigma) = \sum_n \mathcal{J}_n(\a)e^{-in\sigma}$ etc. 
These modes follow the centrally extended interpolating algebra
\begin{subequations}\label{interpol}
\begin{eqnarray}
    &&[\mathcal{J}_n(\a),\mathcal{J}_m(\a)] = (n-m)\mathcal{J}_{n+m}(\a) + \frac{c_L}{12}\,n(n^2-1)\delta_{n+m,0} \\
    &&[\mathcal{J}_n(\a),\mathcal{H}_m(\a)] = (n-m)\mathcal{H}_{n+m}(\a) + \frac{c_M}{12}\,n(n^2-1)\delta_{n+m,0} \\
    &&[\mathcal{H}_n(\a),\mathcal{H}_m(\a)] = \frac{1}{ (\cosh^2{\theta} + \sinh^2{\theta})^2}\big[(n-m)\mathcal{J}_{n+m}(\a) + \frac{c_L}{12}\,n(n^2-1)\delta_{n+m}\big].
\end{eqnarray}
\end{subequations}
Comparing these to a pair of undeformed Virasoro algebras with central charges $c,\bar{c}$, we can easily read off the deformed central terms:
\begin{equation}
    c_L = c-\bar{c}\,,\qquad c_M = (\cosh^2{\theta} + \sinh^2{\theta})^{-1}(c+\bar{c}).
\end{equation}
Observe that what we have got here is an effective contraction of the central extensions as $\theta \to \infty$. Near this extreme value, the scaling factor $ (\cosh^2{\theta} + \sinh^2{\theta})^{-1}$ gradually goes to zero, making sure that we have the centrally extended BMS$_3$ algebra.

\subsubsection{Deformed spectrum}
To calculate the deformed spectrum, we consider the scalars compactified over $D$ circles of radius $R_i$. For the undeformed case, the periodic boundary condition was applied on the fields 
\begin{equation}
    \phi_i (\tau, \sigma + 1) \equiv \phi_i(\tau, \sigma)+ m_i R_i,
\end{equation}
where $m_i$ are the winding numbers and we have rescaled the periodicity for simplicity.
This, in conjunction with the relevant equations of motion, leads to the following mode expansion of the fields
\begin{equation}
	\phi_i (\tau, \sigma)=\phi_i^{(0)}+\frac{n_i}{  R_i}\tau + m_i R_i \sigma+ \frac{i}{\sqrt{2}}\sum_{n \neq 0}\frac{1}{n} \left\{ j^i_n e^{-2\pi{in(\tau + \sigma)}}+\bar{j}^i_n e^{-2\pi{in(\tau -\sigma)}}\right\}.
\end{equation}
The Kaluza-Klein (KK) momentum $(n_i)$ and winding numbers are related to zero modes of the undeformed field as 
\begin{eqnarray}
    m^i = j^i_0 - \bar{j}^i_0\,,\quad n^i = j^i_0 + \bar{j}^i_0.
\end{eqnarray}
Since we still have oscillatory modes in the deformed theory, as made clear by the equations of motion, one can also impose similar periodic boundary conditions in the basis:
\begin{equation}
    \phi_i (\tau, \sigma + 1) \equiv \phi_i(\tau, \sigma)+ \bar{m}_i \bar{R}_i,
\end{equation}
with corresponding new winding numbers $\bar{m}_i $ (and KK momenta $\bar{n}_i$) defined on circles of radii $\bar{R}_i$. 
Reminding ourselves that in the deformed theory the scalar field still follows a wave equation \eqref{waveeqndef} with rescaled speed of light, this leads to the following mode expansion of the fields in terms of oscillators $j(\alpha)$:
\begin{equation}\label{defmode}
	\phi_i (\tau, \sigma;\alpha)=\phi_i^{(0)}+\frac{\bar{n}_i}{  \bar{R}_i}\tau + \bar{m}_i \bar{R}_i \sigma+ \frac{i}{\sqrt{2}}\sum_{n \neq 0}\frac{1}{n} \left\{ j(\alpha)^i_n e^{-2\pi{in(\kappa\tau + \sigma)}}+\bar{j}(\alpha)^i_n e^{-2\pi{in(\kappa\tau -\sigma)}}\right\}.
\end{equation}
Note that when $\kappa\to 0$, the oscillatory nature of the solution gets disturbed and it leads to an expansion linear in $\tau$  when the limit is taken carefully \footnote{See \cite{Bagchi:2015nca} for a detailed discussion on this limit on the mode expansion in the context of null strings. One could find the near-BMS version of our Bogoliubov transformations by taking $\kappa\to 0$ limit on these mode expansions and redefining oscillators.}. 

We now focus on the flow of spectrum. Away from the end point of $\kappa\to 0$, the underlying symmetry algebra is unaltered. Similar to the undeformed case, the winding and KK momenta can still be related to the zero modes of \eqref{defmode}:
\begin{eqnarray}
    \bar{m}^{i} = j(\alpha)^i_0 - \bar{j}(\alpha)^i_0\,,\quad \bar{n}^i = j(\alpha)^i_0 + \bar{j}(\alpha)^i_0.
\end{eqnarray}
Using the explicit Bogoliubov transformations between the two sets of oscillatory modes as defined in \eqref{bogodef}, the relations between two sets of quantities can be expressed as:
\begin{eqnarray}
    \bar{m}^{i} = (\cosh{\theta} - \sinh{\theta})m^i = \kappa m^i,\quad \bar{n}^{i} = (\cosh{\theta} + \sinh{\theta})n^i = \frac{1}{\kappa}n^i,
\end{eqnarray}
with $\kappa$ being the effective speed of light. 
We also assume a highest weight representation for the CFT vacuum to start with, so that for a generic state in the undeformed theory is labeled by weights $h,\bar{h}$:
\be
\mathcal{L}_0|h,\bar{h}\rangle = h|h,\bar{h}\rangle ,~~\mathcal{\bar{L}}_0|h,\bar{h}\rangle = \bar{h}|h,\bar{h}\rangle.
\ee
From the deformed mode expansion of the fields \eqref{defmode} one can easily find the zero modes of the Virasoro generators:
\begin{subequations}
\begin{eqnarray}
    \cL_0(\alpha) &=& \frac{1}{4}\Big(\frac{\bar{n}_i}{ \bar{R}_i} +\bar{m}_i\bar{R}_i\Big)^2 + \frac{1}{2}\sum_{n \neq 0}j(\alpha)^i_{-n}~j(\alpha)^i_{n} \\
    \Bar{\cL}_0(\alpha)&=& \frac{1}{4}\Big(\frac{\bar{n}_i}{ \bar{R}_i} -\bar{m}_i\bar{R}_i\Big)^2 + \frac{1}{2}\sum_{n \neq 0}\bar{j}{(\alpha)}^i_{-n}~\bar{j}{(\alpha)}^i_{n}.
\end{eqnarray}
\end{subequations}
Acting on a deformed state, the conformal dimension is given by the deformed winding and KK momenta:
\begin{subequations}
\begin{eqnarray}
    h(\alpha) = \frac{1}{4}\Big(\frac{\bar{n}_i}{ \bar{R}_i} +\bar{m}_i\bar{R}_i\Big)^2,~~~
    \bar{h}(\alpha) = \frac{1}{4}\Big(\frac{\bar{n}_i}{ \bar{R}_i} -\bar{m}_i\bar{R}_i\Big)^2
\end{eqnarray}
\end{subequations}
Given the above, one can check the combinations
\begin{eqnarray}
    && h(\alpha) - \bar{h}(\alpha) = \bar{n_i}\bar{m_i} = n_i m_i = h-\bar{h}\\
    && h(\alpha) + \bar{h}(\alpha) = \frac{1}{2}\bigg(\left(\frac{\bar{n}_i}{ \bar{R}_i}\right)^2 + (\bar{m_i}\bar{R}_i)^2\bigg) = \frac{1}{2}\bigg(\frac{1}{\kappa^2}\left(\frac{n_i}{ \bar{R}_i}\right)^2 + \kappa^2(m_i\bar{R}_i)^2\bigg).
\end{eqnarray}
Clearly, the spectrum of the theory will be dominated by the KK momenta in the Carroll limit $\kappa\to 0$. {\footnote{This is again in keeping with results from compactified tensionless string literature \cite{Bagchi:2022iqb, Banerjee:2023ekd}, where it was already found that there is no winding number contribution to the mass formula.}} Moreover, one could note that the factors of speed of light in the mass spectrum can be re-absorbed by rescaling the radii of compactification $\bar{R}_i \to \kappa \bar{R}_i$ when $\kappa$ is finite. For $\kappa \to 0$, this implies again, a de facto contraction of the $(h+\bar{h})$ combination. 
\medskip

The take home message from this section is that the diagonal $J\bar{J}$-like deformations dial the speed of light of a field theory, thereby smoothly interpolating between the relativistic theory to the Carrollian theory at the end of the flow. The presence of such deformations implies a mixing of modes in the form of Bogoliubov transformations, and hence the physical vacuum and the spectrum continuously evolve in the space of couplings as the theory approaches the BMS point. This can also be interpreted as a flow in the space of $SL(2,\mathbb{R})$ unitaries, with the BMS point attributed to a singular point in this parameter space \cite{Banerjee:2022ime}. In the next section, we focus on a non-diagonal current-current deformation, which will also induce a spectral flow in the space of couplings, however there would be no continuous change in the physical vacuum and the appearance of BMS will be abrupt. Intriguingly, as we will see, that scenario can also be justified as a singular limit in $SL(2,\mathbb{R})$ transformation space.

\section{The antisymmetric deformation}\label{sec4}

In this section, we turn our attention to the free relativistic scalar theory together with antisymmetric deformations of form $\epsilon_{ab}J^a\bar J^{b}$ that are exactly marginal. A salient feature of such deformations is these mix spacetime indices on the fields, as opposed to the symmetric $J\bar{J}$ case we discussed previously where spacetime indices just decouple from each other. We will see that this feature leads to interesting modifications at the level of boundary conditions. 

\subsection{The classical theory}
Instead of any number of scalar fields like we considered previously, we now have to consider an even number of them. We start with the simplest case, viz. the two dimensional Lagrangian density of two bosonic fields $\phi_{1,2}$ deformed by a $J^1\wedge \bar{J}^2$ operator. Here the currents $J^i = \partial_z\phi^i$ and $\overline{J}^i = \partial_{\bar{z}}\phi^i$ are the holomorphic and antiholomorphic $U(1)$ currents of the theory. \footnote{Addition of this operator can be  equivalently seen as turning on a constant B-field (see Appendix \ref{Bfield}) on the worldsheet of a string moving on a flat Euclidean 2-dimensional target space.} Our notations are in sync with those in \cite{Guica:2021fkv}, where the deformation was defined recursively and the full summed action is:
\be\label{lagwedge}
	\mathcal{L}^{(g)}=\frac{1}{1+g^2}\left[\frac{1}{2} 
	\sum_{i=1,2}\left(\left(\partial_\tau \phi_i\right)^2-\left(\partial_\sigma \phi_i\right)^2\right)
	-g\left(\partial_\tau \phi_1 \partial_\sigma \phi_2-\partial_\sigma \phi_1 \partial_\tau \phi_2\right)\right].
\ee
Starting from the deformed Lagrangian above, we now find momentum densities corresponding to the two fields:
\be\label{conj_mom_phi1phi2}
\Pi_1=\frac{1}{1+g^2}\left({\dot{\phi}_1}-g \phi_2^{\prime}\right), 
\qquad 
\Pi_2=\frac{1}{1+g^2}\left({\dot{\phi}_2}+g \phi_1^{\prime}\right).
\ee
Note that $g$ here is a coupling constant which is not necessarily small, and in fact it will be some upper limit on the value of this coupling that we will be interested in, mirroring our computations in the preceding section. The Hamiltonian density is then obtained in the standard way
\bea
	\mathcal{H}^{(g)}
	=\sum_{i=1,2}\Pi_i \dot{\phi}_i-\cL^{(g)}
	=\frac{1}{2}\left(\left(1+g^2\right)\left(\Pi_1^2+\Pi_2^2\right)+{\phi_1'}^2+{\phi_2'}^2\right)
	+ g\left(\Pi_1 \phi'_2-\Pi_2 \phi'_1\right).
	\label{ham.in.phi12}
\ena
This density can be represented in a matrix form\footnote{To be precise, the off diagonal component is not unique, but here is chosen
by a requirement for the matrix to be symmetric.}
\be\label{hamil.in.phi12}
	\mathcal{H}^{(g)}
	=
	\,\begin{pmatrix}
	\Pi_1 \\ \Pi_2 \\ \phi_1' \\ \phi_2'
	\end{pmatrix}^{\rm T}
	M^{(g)}
	\begin{pmatrix}
	\Pi_1 \\ \Pi_2 \\ \phi_1' \\ \phi_2'
	\end{pmatrix}\ ,
	\qquad \text{where} \qquad
	M^{(g)}=\begin{pmatrix}
	\frac{1+g^2}{2} & 0 & 0 & \frac{g}{2} \\
	0 & \frac{1+g^2}{2} & -\frac{g}{2} & 0 \\
	0 & -\frac{g}{2} & \frac{1}{2} & 0 \\
	\frac{g}{2} & 0 & 0 & \frac{1}{2}
	\end{pmatrix}\ .
\ee
The matrix $M^{(g)}$ has two sets of eigenvalues:
\be
	\lambda^{(g)}_\pm=\frac{2+g^2\pm g\sqrt{g^2+4}}{4}=\frac{\gamma_\pm^2}{2}\ ,
	\qquad\text{where}\qquad
	\gamma_\pm=\frac{\sqrt{g^2+4}\pm g}{2}=\frac{1}{\gamma_\mp}\ .
\ee
Now one similarly calculates the corresponding orthonormal eigenvectors:
\be
	u^{+}_1=c^+
	\begin{pmatrix}
	1 \\ 0 \\ 0 \\ \gamma_-
	\end{pmatrix}
	\quad
	u^{+}_2=c^+
	\begin{pmatrix}
	0 \\ 1 \\ -\gamma_- \\ 0
	\end{pmatrix}
	\quad
	u^{-}_1=c^-
	\begin{pmatrix}
	1 \\ 0 \\ 0 \\ -\gamma_+
	\end{pmatrix}
	\quad
	u^{-}_2=c^-
	\begin{pmatrix}
	0 \\ 1 \\ \gamma_+ \\ 0
	\end{pmatrix}	
\ee
where we have the parameters
\be{}
c^\pm=\frac{1}{\sqrt{1+2\lambda_\mp}},
\ee
and thus $M^{(g)}$ is diagonalised as
\be
	\Lambda={\rm diag}(\lambda^{(g)}_+,\lambda^{(g)}_+,\lambda^{(g)}_-,\lambda^{(g)}_-)
	=S^{\rm T}M^{(g)}S
\ee
by an element in SO(4)
\be
	S=\begin{pmatrix}
	u^{+}_1 & u^{+}_2 & u^{-}_1 & u^{-}_2
	\end{pmatrix}.
\ee
Then the Hamiltonian density can be represented in a simple bilinear form in terms of redefined field variables $\Xi$:
\be\label{hamil.in.Xi}
	\mathcal{H}^{(g)}
	=\,\Xi^{\rm T}\Lambda \Xi
	=\,\lambda^{(g)}_+\left\{\Pi_+^2+(\chi'_-)^2\right\}
	+\,\lambda^{(g)}_-\left\{\Pi_-^2+(\chi'_+)^2\right\}
\ee
where the set of fields $\Xi$ is defined by
\be\label{def.Xi}
	\Xi=
	\begin{pmatrix}
	\Pi_+ \\ \chi'_- \\ -\Pi_- \\ \chi'_+
	\end{pmatrix}
	:=S^{\rm T}
	\begin{pmatrix}
	\Pi_1 \\ \Pi_2 \\ \phi_1' \\ \phi_2'
	\end{pmatrix}
	=
	\begin{pmatrix}
	c^+(\Pi_1+\gamma_-\phi_2') \\
	c^+(\Pi_2-\gamma_-\phi_1') \\
	c^-(\Pi_1-\gamma_+\phi_2') \\
	c^-(\Pi_2+\gamma_+\phi_1')
	\end{pmatrix}
\ee
Note that at $g=0$, the above redefined fields do not reduce to the original variables, but rather to:
\be\label{fieldredef}
	\left.\Xi\right|_{g=0}
	=
	\begin{pmatrix}
	\frac{1}{\sqrt{2}}\left(\dot{\phi}_1+\phi_2'\right) \\
	\frac{1}{\sqrt{2}}\left(\dot{\phi}_2-\phi_1'\right)  \\
	\frac{1}{\sqrt{2}}\left(\dot{\phi}_1-\phi_2'\right)  \\
	\frac{1}{\sqrt{2}}\left(\dot{\phi}_2+\phi_1'\right) 
	\end{pmatrix}.
\ee
This field re-definition turns out to be very useful. As we will see below, these are precisely the combinations that will allow us to track the flow of the theory as the deformation is dialled and will be crucial to see how even at the end point of the flow, the formalism does not break down but there is the emergence of a different spacetime algebra. 

\medskip

Some comments are needed for consistency of this field re-definition. Note that the transformation mixes canonical variables for different fields. This is expected in a theory with an antisymmetric field contribution. This redefinition re-diagonalises the deformed Hamiltonian density. However it is not clear at the outset whether this is a symplectomorphism for the dynamical variables as well. We discuss this point at length below. 

\subsubsection{Dynamics of new conjugate variables}
We now elaborate on the structure of redefined field variables \eqref{fieldredef} introduced above. The dynamics of the deformed Hamiltonian system is still governed by the canonical relation
\be\label{canon.rel}
	\{\phi_i({\sigma}),\Pi_j(\sigma')\}_{\rm PB}=
	-\{\Pi_i(\sigma),\phi_j(\sigma')\}_{\rm PB}=\delta_{ij}\delta(\sigma-\sigma')
\ee
as well as the time evolution equation
\be\label{time.evol}
	\dot{\mathcal{O}}=\{\mathcal{O},H\}_{\rm PB}
	\qquad\quad
	H=\int d\sigma \,\cH^{(g)}
\ee
for any physical observable $\mathcal{O}$.
We may first examine the equations of motion for fields $\phi_i$ as a consistency check for \eqref{canon.rel},
and then, figure out conjugate pairs among components of the field variables in $\Xi$. The time evolutions of variables ${\phi}_i$ are given by
\be\begin{aligned}
	\dot{\phi}_1
	&=\{\phi_1,H\}_{\rm PB}=(1+g^2)\Pi_1+ g \phi'_2, 
	\\
	\dot{\phi}_2
	&=\{\phi_2,H\}_{\rm PB}=(1+g^2)\Pi_2- g \phi'_1.
\end{aligned}\ee
These reproduce the relations of momentum densities with the fields \eqref{conj_mom_phi1phi2}, as expected. Further, the time evolutions of momentum densities are
\be
	\dot{\Pi}_1
	=\{\Pi_1,H\}_{\rm PB}=\phi''_1- g \Pi'_2, 
	\quad \dot{\Pi}_2
	=\{\Pi_1,H\}_{\rm PB}=\phi''_2+ g \Pi'_1.
\ee
Combining the above sets of equations together gives rise to the equations of motion for fields $\phi_i$, which simply turn out to be wave equations: 
\be
	\ddot{\phi}_i-\phi''_i=0.
\ee
Note here that unlike the symmetric deformation, the antisymmetric one does not gradually change the speed of light in the deformed theory. This is clear in hindsight since the deformation term can be regarded as a boundary term, and hence should not change the equations of motion. 

\medskip

Next, we specify conjugate pairs among the new fields, i.e. components of $\Xi$. In order to do that, we work out the Poisson brackets for them using definitions as in \eqref{def.Xi}. This leads to,
\be
	\{\Pi_+(\sigma),\Pi_-(\sigma')\}_{\rm PB}=\{\chi'_+(\sigma),\chi'_-(\sigma')\}_{\rm PB}=0, 
\ee
while
\bea
	\{\Pi_+(\sigma),\chi'_-(\sigma')\}_{\rm PB}
	&=\left(c^+\right)^2\gamma_-\left(
	\pd_{\sigma}+\pd_{\sigma'}
	\right)\delta(\sigma-\sigma')=0, 
	\\
	\{\Pi_+(\sigma),\chi'_+(\sigma')\}_{\rm PB}
	&=c^+c^-\left(\gamma_++\gamma_-\right)\pd_{\sigma}\delta(\sigma-\sigma')
	=\pd_{\sigma}\delta(\sigma-\sigma'). 
\ena
Thus, integrating over $\sigma$, we conclude that
$\chi_+$ and $\Pi_+$ are indeed a pair of conjugate variables that satisfies
\be\label{PBchi+}
	\{\chi_+(\sigma),\Pi_+(\sigma')\}_{\rm PB}=\delta(\sigma-\sigma')
\ee
It is worth emphasising that although we considered $\chi_+$ to be the dynamical variable
and $\Pi_+$ as its conjugate, this is a choice. Another viable choice is $\tilde{\chi}_+=\Pi_+$
as the dynamical variable and $\tilde{\Pi}_+=-\chi_+$ as its conjugate.
This arbitrariness comes from a freedom of choice of Lagrangian submanifold on a symplectic manifold.
For completeness, we also see
\bea
	\{\chi'_-(\sigma),\Pi_-(\sigma')\}_{\rm PB}
	&=\pd_{\sigma}\delta(\sigma-\sigma')
	\\
	\{\Pi_-(\sigma),\chi'_+(\sigma')\}_{\rm PB}
	&=0
\ena
which shows that $\chi_-$ and $\Pi_-$ will indeed be another pair of conjugate variables:
\be\label{PBchi-}
	\{\chi_-(\sigma),\Pi_-(\sigma')\}_{\rm PB}=\delta(\sigma-\sigma').
\ee
Furthermore, since the resulting system is still Hamiltonian, we write the equations of motion:
\be\label{time.evol.chipm}
	\dot{\chi}_+=\frac{\pd\cH^{(g)}}{\pd \Pi_+}=2\,\lambda^{(g)}_+\Pi_+
	\qquad\quad
	\dot{\chi}_-=\frac{\pd\cH^{(g)}}{\pd \Pi_-}=2\,\lambda^{(g)}_-\Pi_-
\ee
which are consistent with the Poisson brackets in \eqref{PBchi+} and \eqref{PBchi-},
as well as the time flow equation \eqref{time.evol}, i.e., 
we also consistently find $\dot{\chi}_\pm=\{{\chi}_\pm,H\}_{\rm PB}=2\,\lambda^{(g)}_\pm\Pi_\pm$.
All of this leads to:
\be\label{time.evol.Pipm}
\begin{aligned}
	\dot{\Pi}_{\pm}&=\left\{{\Pi}_{\pm},H\right\}_{\rm PB}=2\,\lambda^{(g)}_{\mp}\chi''_{\pm}
\end{aligned}\ee
and thus we find that the equation of motion of $\chi_\pm$ is the same wave equation as that of $\phi_{i}\,(i=1,2)$, whatever be the value of the coupling $g$:
\bea
	\ddot{\chi}_+
	&\overset{\eqref{time.evol}}{=}\left\{\dot{\chi}_+,H\right\}_{\rm PB}
	\overset{\eqref{time.evol.chipm}}{=}2\,\lambda^{(g)}_+\left\{\Pi_+,H\right\}_{\rm PB}\nn
	&\overset{!}{=}2\,\lambda^{(g)}_+\dot{\Pi}_+
	\overset{\eqref{time.evol.Pipm}}{=}4\,\lambda^{(g)}_+\lambda^{(g)}_-\chi''_+
	=\chi''_+
\ena
where at $!$ we see that it is the relation obtained by simply taking a differential of \eqref{time.evol.chipm}
with respect to $\tau$, providing a check for self-consistency of our system.
The resultant equations of motion are expected, because in any case $\Pi_\pm$ and $\chi'_\pm$ are linear combinations of
$\{\dot{\phi}_1,\dot{\phi}_2,{\phi}'_1,{\phi}'_2\}$.
\medskip

Before going further, we may convince the reader of the consistency of
the relations in \eqref{time.evol.chipm} with the definition \eqref{def.Xi}. For example, take
the fourth component in the column matrix $\Xi$
\be
	\chi'_+=c^-(\Pi_2+\gamma_+\phi_1'),
\ee
which implies
\bea
	\chi_+(\sigma)
	&=c^-\int^\sigma_{\sigma_0} d\sigma'(\Pi_2+\gamma_+\phi_1')(\sigma')\nn
	&=c^-\frac{g+(1+g^2)\gamma_+}{1+g^2}\phi_1(\sigma)
	+\frac{c^-}{(1+g^2)}\int^\sigma_{\sigma_0} d\sigma'\,\dot{\phi}_2(\sigma')
\label{chip.in.chi12}
\ena
which shows $\chi_+$ is almost a linear combination of $\phi_1$ and $\phi_2$ but involves a integro-differential system. 
Applying to \eqref{chip.in.chi12} a differential with respect to $\tau$, we find
\bea
	\dot{\chi}_+
	=c^-\frac{g+(1+g^2)\gamma_+}{1+g^2}\dot{\phi}_1
	+\frac{c^-}{1+g^2}{\phi}'_2
\label{coins.chip1}
\ena
while, by definition of $\Pi_+$ in \eqref{def.Xi}
\bea
	2\,\lambda^{(g)}_+\Pi_+
	&=2\,\lambda^{(g)}_+c^+(\Pi_1+\gamma_-\phi_2')\nn
	&=\frac{2\lambda^{(g)}_+c^+}{1+g^2}\dot{\phi}_1
	+2\,\lambda^{(g)}_+c^+\frac{-g+(1+g^2)\gamma_-}{1+g^2}{\phi}'_2
\label{coins.chip2}
\ena
After some manipulation, we see that each of the coefficients in front of $\dot{\phi}_1$ and ${\phi}'_2$,
respectively, in equations \eqref{coins.chip1} and \eqref{coins.chip2} are identical, thus
ensuring \eqref{time.evol.chipm}. With this we can conclusively claim that our diagonalized system as 
described by the $\pm$ field variables are completely self consistent as a Hamiltonian system. 
\medskip 

Now why did we need to painstakingly make sure of using the right dynamical variables? We then ask the reader to pay attention to our Hamiltonian $\mathcal{H}^{(g)}$ in the \eqref{hamil.in.Xi}. When we crank up the coupling $g \to \infty$, it is clear that $\lambda_{+}^{(g)} \to \infty$, while $\lambda_{-}^{(g)}\to 0$. Since we have established $\pm$ variables as canonical ones, one can easily see
\be
 \mathcal{H}^{(g\to \infty)}\sim \left\{\Pi_+^2+(\chi'_-)^2\right\}
 \ee
 which clearly commutes with itself at two different spatial points. This is, in fact, Carroll dynamics emerging again in this system {\footnote{One may be tempted to compare this with the case of Non-relativistic string theory \cite{Gomis:2000bd} where the B-field in the sigma model becomes critical.}}
{\footnote{Note that using our earlier wisdom, we seem to be unable to quantify this as either electric or magnetic limit, since two different kind of dynamical variables survive. However this is not an issue as we have mentioned earlier choosing $\tilde{\chi}_+'=\Pi_+$
and $\tilde{\Pi}_+=-\chi_+'$ as conjugate variables in \eqref{hamil.in.Xi} is allowed by the algebra. So one could interpolate between a purely electric limit with only momenta surviving to purely magnetic limit with only spatial gradients surviving, and in all cases the symmetry algebra stays the same. One could also demand $g \to -\infty$, which in turn makes  $\lambda_{-}^{(g)} \to \infty$, while $\lambda_{+}^{(g)}\to 0$, as opposite to the previous case. This could be thought of as a ``dual'' limit on \eqref{hamil.in.Xi}. In the next section we will make this more concrete from the algebraic point of view.}}.
 \medskip
 
We now comment on the generalisation of our construction above. The question is whether, if at all, there is a guarantee that one can always diagonalize a current-current deformed scalar field theory in terms of symplectic field redefinitions. If so, it is likely that we would always be able to dial our parameters in a way to reach a truncated regime of the phase space where Carroll symmetry prevails. The answer to this question is unknown from a classical field theory perspective, as far as we understand. From a case-by-case basis, it turns out this can happen for all marginal deformations borne out of $U(1)$ current bilinears, but we do not have a proof for the same. See Appendix \eqref{appD} for some more details and further comments along this direction.
  

\subsubsection{Virasoro algebra to BMS algebra}
As in the case of the symmetric deformation, there are clear indications that the symmetries of the system change drastically as $g\to \infty$. We now make this concrete. Both redefined fields $\chi_\pm$ are subject to the wave equation and conserved currents are
\be
	\cJ^{\pm}=\dot{\chi}_\pm+{\chi}'_\pm=2\lambda^{(g)}_\pm\Pi_\pm+{\chi}'_\pm,
	\quad 
	\overline{\cJ}^{\pm}=\dot{\chi}_\pm-{\chi}'_\pm=2\lambda^{(g)}_\pm\Pi_\pm-{\chi}'_\pm
\ee
which obey the conservation law $\bar{\pd}\cJ^{\pm}=\pd \overline{\cJ}^{\pm}=0$ with $\pd=\pd_\tau+\pd_\sigma$
and $\bar{\pd}=\pd_\tau-\pd_\sigma$. 
The algebra of these currents is as follows:
\begin{subequations}
\bea
         \left\{\cJ^{+}(\sigma),\cJ^{+}(\sigma')\right\}_{\rm PB}
         &=4\lambda^{(g)}_+\pd_\sigma \delta(\sigma-\sigma'),\\
	\left\{\overline{\cJ}^{+}(\sigma),\overline{\cJ}^{+}(\sigma')\right\}_{\rm PB}
	&=-4\lambda^{(g)}_+\pd_\sigma \delta(\sigma-\sigma'),
	\\
	\left\{\cJ^{-}(\sigma),\cJ^{-}(\sigma')\right\}_{\rm PB}
	&=4\lambda^{(g)}_-\pd_\sigma \delta(\sigma-\sigma'),
\\	\left\{\overline{\cJ}^{-}(\sigma),\overline{\cJ}^{-}(\sigma')\right\}_{\rm PB}
	&=-4\lambda^{(g)}_-\pd_\sigma \delta(\sigma-\sigma'),
\ena
\end{subequations}
and all other brackets vanish. 
To make sure these currents have the standard Kac-Moody commutation relation, we normalize them
\be
	J^\pm=\frac{\cJ^{\pm}}{\sqrt{4\lambda^{(g)}_\pm}},
	\qquad\quad
	\overline{J}^\pm=\frac{\overline{\cJ}^{\pm}}{\sqrt{4\lambda^{(g)}_\pm}}.
\ee
Now, applying the Sugawara construction we can again obtain an independent pair of stress tensors
\be
\begin{aligned}
	T(\sigma)&=\frac{1}{2}\left\{J^+(\sigma)J^+(\sigma)+J^-(\sigma)J^-(\sigma)\right\},\\
	\overline{T}(\sigma)&=\frac{1}{2}\left\{\overline{J}^+(\sigma)\overline{J}^+(\sigma)+\overline{J}^-(\sigma)\overline{J}^-(\sigma)\right\}.
\end{aligned}
\ee
As a consistency check, we examine $T+\overline{T}$:
\bea
	T+\overline{T}
	&=\frac{1}{2}\left\{J^+J^++J^-J^-+\overline{J}^+\overline{J}^++\overline{J}^-\overline{J}^-\right\}\nn
	&=\lambda^{(g)}_+\left\{\Pi_+^2+\left({\chi}'_-\right)^2\right\}
	+\lambda^{(g)}_-\left\{\Pi_-^2+\left({\chi}'_+\right)^2\right\}
\ena
which reproduces the Hamiltonian density \eqref{hamil.in.Xi} as expected, while the other orthogonal operator is expressed in a form
\bea
	T-\overline{T}
	&=\frac{1}{2}\left\{J^+J^++J^-J^-+\overline{J}^+\overline{J}^++\overline{J}^-\overline{J}^-\right\}
	=\Pi_+{\chi}'_++\Pi_-{\chi}'_-.
\ena
Let us introduce the generators in a way with suitable scalings, such that the $g\to\infty$ limit can be taken consistently:
\be\label{contract}
	L^{(g)}=T-\overline{T}
	\qquad\qquad
	M^{(g)}=\frac{1}{\lambda^{(g)}_+}(T+\overline{T}).
\ee
It is easy to see the algebra of $L^{(g)}$ is just the Virasoro algebra:
\bea
	\left\{L^{(g)}(\sigma),L^{(g)}(\sigma')\right\}_{\rm PB}
	=2L^{(g)}(\sigma)\pd_\sigma\delta(\sigma-\sigma')+\left(\pd_\sigma L^{(g)}(\sigma)\right) \delta(\sigma-\sigma').
\ena
In a similar manner we find
\bea
	\left\{L^{(g)}(\sigma),M^{(g)}(\sigma')\right\}_{\rm PB}
	&=2M^{(g)}(\sigma)\pd_\sigma\delta(\sigma-\sigma')+\left(\pd_\sigma M^{(g)}(\sigma)\right) \delta(\sigma-\sigma')
	\\
	\left\{M^{(g)}(\sigma),M^{(g)}(\sigma')\right\}_{\rm PB}
	&=\frac{1}{(\lambda^{(g)}_+)^2}
	\left[2L^{(g)}(\sigma)\pd_\sigma\delta(\sigma-\sigma')+\left(\pd_\sigma L^{(g)}(\sigma)\right) \delta(\sigma-\sigma')\right]\nonumber
	\\
	& \overset{g\to\infty}{\longrightarrow}\ 0
\ena
which manifestly shows the transition from a pair of Virasoro algebras to BMS$_3$ algebra as $g$ is dialed up to infinity.

\medskip

We emphasise, that although these equations are similar to \eqref{flowalgebra}, we have arrived at the Carrollian sector in a path very different from the previous symmetric deformation. Unlike the $J. J$ deformation, in this case there was no effective dialling of the speed of light, equations of motion were unchanged and most importantly the generators in \eqref{genbog} and \eqref{contract}, especially the second equalities are completely different. In fact, when written in terms of modes, \eqref{contract} implies a In\"on\"u-Wigner contraction of the two copies of Virasoro algebra (see \eqref{IW}), while the second equation of \eqref{genbog} is clearly something different. 

\medskip

Note that since $\lambda^{(g \to \infty)}\sim g^2$, the contractions in \eqref{contract} as read using the modes \eqref{Lmodes} for very large values of $g$ will be written as: $L^{(\infty)}_n = \mathcal{L}_n - {\mathcal{\bar{L}}_{-n}}$ and $M^{(\infty)}_n = \frac{1}{g^2}\left(\mathcal{L}_n +{\mathcal{\bar{L}}_{-n}}\right)$, which one should again compare with the nature of \eqref{IW}. This standard form of BMS$_3$ generators will be important later on when we go ahead and look at the fate of the spectrum as one goes to $g \to \infty$.

\subsection{Quantization and details of the flow}
We will now move towards a quantization of our deformed system. For the present case, we will consider compactification on circles for both the scalar fields \footnote{In terms of string theory, the above condition is akin to considering compactified boundary conditions in the target space in order to make sure the effect of B-field is visible on the spectrum.}. Let us define the periodic boundary conditions on the original fields as:
\be
	\phi_i\sim \phi_i+ m_i R_i.
	\qquad\quad(i=1,2)
\ee
Here $R_i$ are the compactification radii, associated to which we have Kaluza-Klein (KK) momenta $n_i$ and winding modes $m_i$:
\be
	p_i=\frac{n_i}{R_i}
	\qquad\quad
	w_i=R_i\,m_i
	\qquad\quad
	 n_i,m_i\in\mathbb{Z}.
\ee
We have also rescaled the periodicity in this case $\sigma \in [0,1]$. The deformation term acts as a boundary term and we will need to be very careful about boundary conditions. This is what we address next. 

\subsubsection{Boundary conditions}
We now carefully implement boundary conditions while quantizing the system. Consider a variation over fields $\phi_i\to\phi_i+\delta\phi_i$. Under this, the Lagrangian density in \eqref{lagwedge} varies as
\bea
	\delta \cL^{(g)}
	&=\frac{1}{1+g^2}\left[
	\sum_{i=1,2}\left(\partial_\tau\left[\delta\phi_i\partial_\tau \phi_i\right]
	-\partial_\sigma\left[\delta\phi_i\partial_\sigma \phi_i\right]\right)
	+\overbrace{\left(\partial^2_\sigma \phi_i-\partial^2_\tau \phi_i\right)}^{\text{e.o.m.}} \delta\phi_i  \right]\nn
	&\qquad
	-\frac{g}{1+g^2}\left(\partial_\tau\left[\delta\phi_1 \partial_\sigma \phi_2-\delta\phi_2\partial_\sigma \phi_1 \right]
	+\partial_\sigma\left[\delta\phi_2\partial_\tau \phi_1 -\delta\phi_1 \partial_\tau \phi_2\right]\right).
\ena
The spatial derivatives leave boundary terms after integration:
\bea
	\int_0^{1} d\sigma\,\delta \cL^{(g)}
	&=\frac{1}{1+g^2}\left[
	- \delta\phi_1\left(\partial_\sigma \phi_1-g\,\partial_\tau \phi_2\right)
	- \delta\phi_2\left(\partial_\sigma \phi_2+g\,\partial_\tau \phi_1\right) \right]_{\sigma=0}^{1}
\ena
vanishing of which requires the following boundary conditions:
\begin{subequations}
\bea
	&& \left.\delta\phi_1\left[\partial_\sigma \phi_1-g\,\partial_\tau \phi_2\right]\right|_{\sigma=1}
	 - \left.\delta\phi_1\left[\partial_\sigma \phi_1-g\,\partial_\tau \phi_2\right]\right|_{\sigma=0}=0. \\
	&&  \left.\delta\phi_2\left[\partial_\sigma \phi_2+g\,\partial_\tau \phi_1\right]\right|_{\sigma=1}
	 - \left.\delta\phi_2\left[\partial_\sigma \phi_2+g\,\partial_\tau \phi_1\right]\right|_{\sigma=0}=0.
\ena
\end{subequations}
Now for clarity, let's see what these conditions imply:
\begin{itemize}
\item[(i)] {\bf{Undeformed theory $g=0$}}: For the undeformed theory, the condition are simply:
\be
	 \left.\delta\phi_i \partial_\sigma \phi_i\right|_{\sigma=1}
	 - \left.\delta\phi_i\partial_\sigma \phi_i\right|_{\sigma=0}=0.
\ee
One can use Neumann  ($\left.\partial_\sigma \phi_i\right|_{\sigma=0,1}=0$)  or Dirichlet ($\left.\delta\phi_i\right|_{\sigma=0,1}=0$) boundary conditions to solve this. This is similar to the analysis for the open string.  

One can also employ periodic boundary conditions
($\phi_i(\tau,\sigma+1)=\phi_i(\tau,\sigma)$), however up to
windings, ($\phi_i(\tau,\sigma+1)-\phi_i(\tau,\sigma)= R_i m_i,\ m_i\in\mathbb{Z})$. This is reminiscent of the analysis in closed string theory.

\item[(ii) ]  {\bf{Deformed theory $g\neq0$}}: Since the periodic boundary condition with windings can be written as
\be
	\int_{\sigma}^{\sigma+1} d\sigma'\, \pd_{\sigma'}\phi_i(\tau,\sigma')= R_i m_i,
\ee
a natural extension of this condition in the presence of the coupling $g$ is clearly of the following form:
\be\label{redef-period}
	\int_{\sigma}^{\sigma+1} d\sigma'\, \pd_{\sigma'}\widetilde{\phi}_i
	:=\int_{\sigma}^{\sigma+1} d\sigma'\, \left(\pd_{\sigma'}\phi_i-g\epsilon_{ij}\pd_{\tau}\phi_j\right)(\tau,\sigma')
	=: \gamma_i 
\ee
where we use the tilde to distinguish a deformed field and $\gamma_i $ is an effective winding number for the theory (see also \cite{Apolo:2019zai}). This in turn implies the boundary condition when written in terms of the undeformed quantum numbers:
\begin{equation}\label{redef-bc}
	\widetilde{\phi}_i (\tau, \sigma + 1)- \widetilde{\phi}_i(\tau, \sigma)
	=  m_iR_i- g\epsilon_{ij}\frac{n_j}{R_j}, 
\end{equation}
which means the effective deformed winding is a combination of undeformed winding and momenta, albeit coming from different fields. 
\end{itemize}

\subsubsection{Equations of motion}
It is straightforward to see that the deformed fields will still satisfy wave equations of motion. 
Considering the relation \eqref{redef-period} should be valid for any $\sigma$ as the origin of the periodic integral, implies that the integrands on both sides should be equated 
\be\label{redef-field-current}
	\pd_{\sigma}\widetilde{\phi}_i
	=\pd_{\sigma}\phi_i-g\epsilon_{ij}\pd_{\tau}\phi_j\ .
\ee
Note that this is a very non-local boundary condition that links the different fields and their temporal and spatial derivatives.
This gives rise to relations of the dynamical variables $\widetilde{\phi}_i$ constructed in terms of $\phi_i$:
\be\label{redef-tilde-old}
	\widetilde{\phi}_i(\tau,\sigma)
	=\phi_i(\tau,\sigma)
	-g\epsilon_{ij}\int_{\sigma_0}^\sigma d\sigma' \,\pd_{\tau}\phi_j(\tau,\sigma')+F_i(\tau,\sigma_0)
\ee
where $F_i$ are $\sigma$-independent functions
\be
	F_i(\tau,\sigma_0)=\widetilde{\phi}_i(\tau,\sigma_0)-\phi_i(\tau,\sigma_0).
\ee

Now since we are dealing with just a boundary term, the equations of motion coming from this action should not change, i.e. 
the redefined field $\widetilde{\phi}_i$ obeys the same equation of motion as $\phi_i$ as mentioned before,
\bea
	(\partial^2_\sigma-\partial^2_\tau)\widetilde{\phi}_i
	&=(\partial^2_\sigma-\partial^2_\tau){\phi}_i
	-g\epsilon_{ij}(\partial^2_\sigma-\partial^2_\tau)\int_{\sigma_0}^\sigma d\sigma' \,\pd_{\tau}\phi_j(\tau,\sigma')
	-\pd_{\tau}^2F_i\nn
	&=-g\epsilon_{ij}\int^\sigma d\sigma' \,\pd_{\tau}(\partial^2_{\sigma'}-\partial^2_\tau)\phi_j(\tau,\sigma')
	=0
\ena
by simply using the equations of motion for $\phi_i$\footnote{Note that this mandates the unknown function to be at most linear in $\tau$: $F_i=F_i^{(0)}+\tau F_i^{(1)}$.}.
Hence, $\widetilde{\phi}_i$ will have a Fourier mode expansion, however with the boundary condition \eqref{redef-bc} taken care of,
which is solved by
\begin{equation}
\begin{aligned}
	\widetilde{\phi}_i (\tau, \sigma)&=\widetilde{\phi}_i^{(0)}
	+\widetilde{\phi}_i^{(1)}\tau + \left(m_i R_i - g\epsilon_{ij}\frac{n_j}{R_j} \right)\sigma
	+\sum_{n \neq 0}\frac{1}{n} \left\{ \widetilde{\alpha}^i_n e^{-2\pi{in(\tau + \sigma)}}
	+\widetilde{\widetilde{\alpha}}^i_n e^{-2\pi{in(\tau -\sigma)}}\right\}
\end{aligned}
\end{equation}
One can see the linear piece in $\tau$, in this case $\widetilde{\phi}_i^{(1)}$ can be checked for self-consistency using boundary conditions,
noting that \eqref{redef-field-current} implies:
\be
	0=\pd_{\sigma}\left(\pd_{\sigma}\widetilde{\phi}_i
	-\pd_{\sigma}\phi_i+g\epsilon_{ij}\pd_{\tau}\phi_j\right)
	\overset{\rm e.o.m.}=\pd_{\tau}\left(\pd_{\tau}\widetilde{\phi}_i
	-\pd_{\tau}\phi_i+g\epsilon_{ij}\pd_{\sigma}\phi_j\right)
\ee
as well as
\be
	0=\pd_{\tau}\left(\pd_{\sigma}\widetilde{\phi}_i
	-\pd_{\sigma}\phi_i+g\epsilon_{ij}\pd_{\tau}\phi_j\right)
	\overset{\rm e.o.m.}=\pd_{\sigma}\left(\pd_{\tau}\widetilde{\phi}_i
	-\pd_{\tau}\phi_i+g\epsilon_{ij}\pd_{\sigma}\phi_j\right)
\ee
thus one could easily find a solution where,
\be
	\pd_{\tau}\widetilde{\phi}_i-\pd_{\tau}\phi_i+g\epsilon_{ij}\pd_{\sigma}\phi_j,
	=C_i \,(\rm const.)
\ee
which can be heuristically chosen as $C_i=g\epsilon_{ij}R_jn_j$, resulting in
\bea
	\widetilde{\phi}_i^{(1)}
	=\int_0^{1} d\sigma\,\pd_{\tau}\widetilde{\phi}_i(\tau,\sigma)
	=\frac{n_i}{R_i},
\ena
which one readily expects for zero modes since only the winding modes are affected by the new boundary conditions.
All of this leads to the following  final mode expansion of our deformed fields:
\begin{equation}
	\phi_i (\tau, \sigma)=\phi_i^{(0)}+\frac{n_i}{  R_i}\tau + (m_i R_i - g\epsilon_{ij}\frac{n_j}{R_j} )\sigma+\sum_{n \neq 0}\frac{1}{n} \left\{ \alpha^i_n e^{-{2\pi in(\tau + \sigma)}}+\tilde{\alpha}^i_n e^{-{2\pi in(\tau -\sigma)}}\right\}
\end{equation}
where we have now removed the extra tildes to simplify the notation. This mode expansion for the fields will be used below.

\subsubsection{Deformed generators}
We make the important observation that although the zero modes of the expansion for the field have now changed, the oscillator modes are unchanged.  Comparing with generic conformal field theory mode expansion for a compactified target space \footnote{See \cite{DiFrancesco:1997nk} for example.}, we can define the shifted (or ``flowed") modes:
\begin{eqnarray}\label{flowzeroalpha}
	\alpha^i_{n} \rightarrow  \alpha^i_{n} - \frac{g}{\sqrt{2}}\epsilon_{ij}\frac{n_j}{R_j}\delta_{n,0}, \quad \tilde{\alpha}^i_{n} \rightarrow \tilde{\alpha}^i_{n} + \frac{g}{\sqrt{2}}\epsilon_{ij}\frac{n_j}{R_j}\delta_{n,0}
\end{eqnarray}
which keep KK momenta same, while deforming the periodicity. Such shifted modes do not change the mode algebra or the definition of the vacuum, which remains the highest weight vacuum:
\be{}
\alpha_n |0\rangle = 0 = \tilde\alpha_n|0\rangle, ~~~~~ \text{for} \, \, n> 0.
\ee
We emphasise that this is markedly different from the $J\overline{J}$ case, where the vacuum changed via a one parameter Bogoliubov transformation throughout the flow \eqref{bogovac}.
 From the mode expansion, we can construct usual generators:
\begin{eqnarray}
    &&\cL_n = \frac{1}{2}\sum_{p\in \mathbb{Z}}\alpha^i_{n-p}\alpha^i_{p} \quad \text{for} \, \, n \neq 0, \, \, \text{and} \quad \cL_0 = \frac{1}{2}\sum_{p \neq 0}\alpha^i_{-p}\alpha^i_{p} + \frac{1}{2}\alpha^i_{0}\alpha^i_{0},
\end{eqnarray}
which extends similarly to the antiholomorphic sector. 
Now using shifted zero modes, we can easily get the expression for deformed $\cL_0$ and $\Bar{\cL}_0$ 
\begin{subequations}
\begin{eqnarray}
    \cL_0^{(g)} &=& \frac{1}{4}\Big(\frac{n_i}{R_i} + m_i R_i - g\epsilon_{ij}\frac{n_j}{R_j} \Big)^2 + \frac{1}{2}\sum_{n \neq 0}\alpha^i_{-n}\alpha^i_{n}, \\
    \Bar{\cL}_0^{(g)} &=& \frac{1}{4}\Big(\frac{n_i}{R_i} - m_i R_i + g\epsilon_{ij}\frac{n_j}{R_j} \Big)^2 + \frac{1}{2}\sum_{n \neq 0}\tilde{\alpha}^i_{-n}\tilde{\alpha}^i_{n}.
\end{eqnarray}
\end{subequations}
In the above, we use the $(g)$ superscript to identify the ``flow"-ed quantities.
Similarly, the shifted conformal dimension is given by action of Virasoro zero modes on a primary state $|h^{(g)},\Bar{h}^{(g)}\rangle$:
\be
\cL_0^{(g)} |h^{(g)},\Bar{h}^{(g)}\rangle = h^{(g)} |h^{(g)},\Bar{h}^{(g)}\rangle,  \quad \Bar{\cL}_0^{(g)} |h^{(g)},\Bar{h}^{(g)}\rangle = \Bar{h}^{(g)} |h^{(g)},\Bar{h}^{(g)}\rangle, 
\ee
where the explicit shifted weights are
\begin{subequations}
\begin{eqnarray}
    h^{(g)} &=& \frac{1}{4}\Big(\frac{n_1}{R_1} + m_1 R_1 - g\frac{n_2}{R_2} \Big)^2 + \frac{1}{4}\Big(\frac{n_2}{R_2} + m_2 R_2 + g\frac{n_1}{R_1} \Big)^2\\
    \Bar{h}^{(g)} &=& \frac{1}{4}\Big(\frac{n_1}{R_1} - m_1 R_1 + g\frac{n_2}{R_2} \Big)^2 + \frac{1}{4}\Big(\frac{n_2}{R_2} - m_2 R_2 - g\frac{n_1}{R_1} \Big)^2.
\end{eqnarray}
\end{subequations}
We would be interested in the combinations
\begin{subequations}
\begin{eqnarray}
    && h^{(g)} - \Bar{h}^{(g)} =h-\Bar{h}= n_im_i, \\
    && h^{(g)} + \Bar{h}^{(g)} = \Bigg[\frac{1}{2}\Big((1+g^2)\frac{n_i^2}{R_i^2} + m_i^2 R_i^2\Big) +g\Big(\frac{R_2}{R_1}n_1m_2 - \frac{R_1}{R_2}n_2m_1 \Big)\Bigg]. 
\end{eqnarray}\end{subequations}
We see that $h^{(g)} - \Bar{h}^{(g)}$ does not flow, while the other linear combination $h^{(g)} + \Bar{h}^{(g)}$ does. 
A similar computation with shifted zero modes leads to expressions for flowed $\cL_n, \bar{\cL}_n$ with $n \neq 0$:
\begin{subequations}
\begin{eqnarray}
	\cL_n^{(g)} &=& \frac{1}{2}\sum_{\substack{p \neq 0 \\ p\neq n}}\alpha^i_{n-p}\alpha^i_{p} +\frac{1}{\sqrt{2}}\Big(\frac{n_i}{R_i} + m_i R_i - g\epsilon_{ij}\frac{n_j}{R_j} \Big) \alpha^i_{n}, \\
	\bar{\cL}_n^{(g)}&=& \frac{1}{2}\sum_{\substack{p \neq 0 \\ p\neq n}}\tilde{\alpha}^i_{n-p}\tilde{\alpha}^i_{p} + \frac{1}{\sqrt{2}}\Big(\frac{n_i}{R_i} - m_i R_i + g\epsilon_{ij}\frac{n_j}{R_j} \Big) \tilde{\alpha}^i_{n}.
\end{eqnarray}\end{subequations}
Combining everything, we can see the Virasoro generators $\cL_n$ and $\bar{\cL}_n$ flows as
\begin{subequations}\label{flowofL}
\begin{eqnarray}
    \cL_n &\rightarrow & 
   \cL_n - \frac{g}{\sqrt{2}}\epsilon_{ij}\frac{n_j}{R_j}\alpha^i_{n} + \frac{g^2}{4}\frac{n_i^2}{R_i^2}\delta_{n,0}\\
    \bar{\cL}_n &\rightarrow & \bar{\cL}_n + \frac{g}{\sqrt{2}}\epsilon_{ij}\frac{n_j}{R_j}\tilde{\alpha}^i_{n} + \frac{g^2}{4}\frac{n_i^2}{R_i^2}\delta_{n,0}.
\end{eqnarray}\end{subequations}
Note that despite the shifted nature, all the generators here still correspond to the conformal algebra for all values of finite $g$. This is of course expected since the deformation that we started out with is an exactly marginal deformation and although the spectrum of the theory changes, the theory still remains conformal. 

\medskip

We can compare this with the usual spectral flow for WZW models \footnote{See \cite{Maldacena:2000hw} and Appendix \eqref{appwzw} for a brief introduction to such spectral flows.}. In this case the spectral flow parameter is the KK momentum associated to the compactified theory. A similar result has also been derived in \cite{Apolo:2019zai} by considering TsT transformations in a AdS$_3$ target space that induce current-current deformations on a string worldsheet.  In what follows, we will go a bit further and show that in the large $g$ limit the flowed algebra in this case conveniently rearranges itself in the form of a BMS invariant theory. 

\subsubsection{$g \to \infty$ limit: Emergent BMS}
We now wish to go to the edge of the parameter space by dialling the coupling $g$. We know from the classical analysis, that we expect a change of the spacetime algebra from relativistic to Carrollian conformal symmetries. We will now see this explicitly in the quantum mechanical theory. Unlike the classical analysis, now that we have interpreted the quantum mechanical deformation first as a change of boundary condition and then as a spectral flow, we now will see that there exists an extreme limit of these boundary conditions or the spectral flow which deforms the relativistic symmetries to Carrollian. We should emphasise that these are completely novel observations. The fact that changing boundary conditions drastically could significantly change a theory is perhaps expected, but changing relativistic symmetries to non-relativistic (Carrollian in this case) by a continuous deformation of boundary condition is remarkable. We now explicitly show this. 

\medskip

To take the $g \to \infty$ limit on the results of the last section, especially of \eqref{flowofL}, let us define a new convenient  basis of generators of the form $ \mathcal{J}_n = {\cL}_n - \bar{{\cL}}_{-n}$  and $\mathcal{H}_n = \frac{1}{g^2}({\cL}_n + \bar{{\cL}}_{-n})$, and observe that they also flow as in the following way:
\begin{eqnarray}
    \mathcal{J}^{(g)}_n &&= {\cL}^{(g)}_n - \bar{{\cL}}^{(g)}_{-n} = {\cL}^{(g=0)}_n - \bar{{\cL}}^{(g=0)}_{-n} - \frac{g}{\sqrt{2}}\epsilon_{ij}\frac{n_j}{R_j}(\alpha^i_{n} + \tilde{\alpha}^i_{-n})\nonumber\\
    \mathcal{H}^{(g)}_n &&= \frac{1}{g^2}({\cL}^{(g)}_n + \bar{{\cL}}^{(g)}_{-n}) = \frac{1}{g^2}({\cL}^{(g=0)}_n + \bar{{\cL}}^{(g=0)}_{-n}) - \frac{1}{\sqrt{2}g}\epsilon_{ij}\frac{n_j}{R_j}(\alpha^i_{n} - \tilde{\alpha}^i_{-n}) + \frac{1}{2}\frac{n_i^2}{R_i^2}\delta_{n,0}\nonumber
\end{eqnarray}
So, we have
\begin{eqnarray}\label{genflow1}
\mathcal{J}^{(g)}_n = \mathcal{\overline{J}}_n -\frac{1}{\sqrt{2}g}\epsilon_{ij}\frac{n_j}{R_j}B^i_n, \quad  
\mathcal{H}^{(g)}_n =  \mathcal{\overline{H}}_n -\frac{1}{\sqrt{2}g}\epsilon_{ij}\frac{n_j}{R_j}A^i_n + {\frac{1}{2}\frac{n_i^2}{R_i^2}\delta_{n,0}}
\end{eqnarray}
where we have redefined:
\be
\alpha^i_{n} - \tilde{\alpha}^i_{-n} = A^i_n, \quad \alpha^i_{n} + \tilde{\alpha}^i_{-n} = \frac{1}{g^2}B^i_n
\ee 
as the new combination of oscillator modes and the superscript $(g=0)$ emphasizes the undeformed generators, so that $\mathcal{\overline{J}}_n=\mathcal{J}^{(g=0)}_n$ and $ \mathcal{\overline{H}}_n = \frac{1}{g^2}\mathcal{H}^{(g=0)}_n$ are the non-flowing parts of the deformed generators. Note that the definition of the stress tensor modes $\mathcal{J}^{(g)}_n$ and $ \mathcal{H}^{(g)}_n$ are commensurate with \eqref{contract} in large $g$ limit. 
Further notice that under the flow of zero modes as in \eqref{flowzeroalpha}, $A^i_0, B^i_0$ also shift in the following way:
\be\label{ABflow}
A_0^i \rightarrow A_0^i - \sqrt{2}g\epsilon_{ij}\frac{n_j}{R_j}\delta_{n,0}, ~~~~B_0^i \rightarrow B_0^i.
\ee
This immediately reminds us of the spectral flow in 3D flat spacetime as described in \cite{Basu:2017aqn} for a holographic setting. In fact one could easily identify \eqref{genflow1} as an automorphism of a BMS$-U(1)$ theory, where the parent Virasoro-Kac-Moody algebras had the same level. In that case $A,B$ are the generators of the affine current algebra.
Now focussing on the deformed Kac-Moody algebra between $\mathcal{H},\mathcal{J}, A,B$, we observe: 
\begin{eqnarray}\label{NLKM1}
    \left[\mathcal{H}^{(g)}_m,A^k_n\right]
    &&= \left[\mathcal{\overline{H}}_m,A^k_n\right] \nonumber\\
  \left[\mathcal{H}^{(g)}_m,B^k_n\right] 
    &&= \left[\mathcal{\overline{H}}_m,B^k_n\right] - \sqrt{2}{g}\epsilon_{ij}\frac{n_j}{R_j}m\delta^{ik}\delta_{m+n,0}\nonumber\\
\left[\mathcal{J}^{(g)}_m,A^k_n\right] 
    &&= \left[\mathcal{\overline{J}}_m,A^k_n\right] - \sqrt{2}{g}\epsilon_{ij}\frac{n_j}{R_j}m\delta^{ik}\delta_{m+n,0}\nonumber\\
     \left[\mathcal{J}^{(g)}_m,B^k_n\right] 
    &&= \left[\mathcal{\overline{J}}_m,B^k_n\right] 
\end{eqnarray}
Also, we can easily work out the Kac-Moody algebra involving  $\mathcal{\overline{J}}_n$ and $\mathcal{\overline{H}}_n$ separately, 
\begin{eqnarray}\label{NLKM2}
    && \left[\mathcal{\overline{H}}_m,B^k_n\right]= \left[{\cL}^{(g=0)}_m + \bar{{\cL}}^{(g=0)}_{-m},\alpha^k_{n} + \tilde{\alpha}^k_{-n}\right] 
     = -nA^k_{m+n}, \nonumber \\
    &&\left[\mathcal{\overline{H}}_m,A^k_n\right] = \frac{1}{g^2}\left[{\cL}^{(g=0)}_m + \bar{{\cL}}^{(g=0)}_{-m},\alpha^k_{n} - \tilde{\alpha}^k_{-n}\right]
    = -\frac{1}{g^4}nB^k_{m+n}, \nonumber\\
    &&\left[\mathcal{\overline{J}}_m,B^k_n\right] = g^2\left[{\cL}^{(g=0)}_m - \bar{{\cL}}^{(g=0)}_{-m},\alpha^k_{n} + \tilde{\alpha}^k_{-n}\right] 
    = -nB^k_{m+n}, \nonumber \\
    &&\left[\mathcal{\overline{J}}_m,A^k_n\right]= \left[{\cL}^{(g=0)}_m - \bar{{\cL}}^{(g=0)}_{-m},\alpha^k_{n} - \tilde{\alpha}^k_{-n}\right] 
    = -nA^k_{m+n}. 
\end{eqnarray}
Note that the flow of zero modes do not affect these commutation relations, and moreover the flowed terms on the RHS of \eqref{NLKM1} are simply cancelled by the flow terms in $A_0, B_0$ coming from the undeformed algebra as defined in \eqref{ABflow}.
Collecting the results from the above computation and taking the $g\to \infty$ limit, everything falls into place and we finally reach at the following Kac-Moody algebra:
\begin{eqnarray}
    \left[\mathcal{H}^{(\infty)}_m,A^k_n\right] &=& 0, \quad \left[\mathcal{H}^{(\infty)}_m,B^k_n\right] = -nA^k_{m+n},     \nonumber\\
    \left[\mathcal{J}^{(\infty)}_m,A^k_n\right] &=& -nA^k_{m+n}, \quad  \left[\mathcal{J}^{(\infty)}_m,B^k_n\right] = -nB^k_{m+n}.
\end{eqnarray}
This is the BMS$-U(1$) Kac-Moody algebra as found in \cite{Bagchi:2023dzx}. Note that unlike in that work, we did not arrive here by contractions of the seed algebra, but by just deforming the generators with $g$ to the extreme value.
\medskip

Coming back to the stress tensor modes, we can easily see the barred generators follow the algebra
\begin{eqnarray}
    &&\left[\mathcal{\overline{J}}_n,\mathcal{\overline{J}}_m\right] = (n-m)\mathcal{\overline{J}}_{m+n}\\
    &&\left[\mathcal{\overline{J}}_n,\mathcal{\overline{H}}_m\right] = (n-m)\mathcal{\overline{H}}_{m+n}\\
    &&\left[\mathcal{\overline{H}}_n,\mathcal{\overline{H}}_m\right] = (n-m)\mathcal{\overline{J}}_{m+n}\frac{1}{g^4} \xrightarrow{g\rightarrow \infty} 0
\end{eqnarray}
This would have easily served as a cousin of the interpolating algebras we have seen before, going from two copies of Virasoro algebra at finite $g$ to BMS$_3$ in infinite $g$. However there is a caveat, since the $g=0$ point is not simply connected in this flow. This makes sense as our basis itself is designed to work in the large $g$ limit so that we can smoothly go up to \eqref{contract}.  Using the above results, the full flowed generators can be found to obey: 
\begin{align}
    \left[\mathcal{J}^{(g)}_n,\mathcal{J}^{(g)}_m\right] 
    &= (n-m)\mathcal{\overline{J}}_{m+n} + \frac{1}{\sqrt{2}g}\epsilon_{ij}\frac{n_j}{R_j}mB^i_{m+n} - \frac{1}{\sqrt{2}g}\epsilon_{ij}\frac{n_j}{R_j}nB^i_{m+n} \nonumber\\
    &= (n-m)\mathcal{J}^{(g)}_{m+n}\nonumber\\
   \left[\mathcal{J}^{(g)}_n,\mathcal{H}^{(g)}_m\right] 
    &= (n-m)\mathcal{\overline{H}}_{m+n} - \frac{1}{\sqrt{2}g}\epsilon_{ij}\frac{n_j}{R_j}mA^i_{m+n} +  \frac{1}{\sqrt{2}g}\epsilon_{ij}\frac{n_j}{R_j}nA^i_{m+n} + n\delta_{n+m,0}\frac{n_i^2}{R_i^2}\nonumber\\
    &= (n-m)\mathcal{H}^{(g)}_{m+n} \nonumber\\
\left[\mathcal{H}^{(g)}_n,\mathcal{H}^{(g)}_m\right] 
    &= \bigg((n-m)\mathcal{\overline{J}}_{m+n} + \frac{1}{\sqrt{2}g}\epsilon_{ij}\frac{n_j}{R_j}mB^i_{m+n} - \frac{1}{\sqrt{2}g}\epsilon_{ij}\frac{n_j}{R_j}nB^i_{m+n}\bigg)\frac{1}{g^4} \nonumber\\
    &= (n-m)\mathcal{J}^{(g)}_{m+n}\frac{1}{g^4}\xrightarrow{g\rightarrow \infty} 0
\end{align}
So in the strict $g\to \infty$ limit, we have obtained the BMS$_3$ or the 2d Conformal Carroll algebra as promised. Since this is the centerless version, we should also mention same logic applies to the central charges of the theory, although we can confirm from \eqref{flowofL} that flowed Virasoro generators satisfy the algebra with same central charges, the consistent $g\to \infty$ limit readily leads to central charges of the flowed BMS algebra:
\be
c_{\cJ} = c-\bar{c},~~~c_{\cH} =\frac{1}{g^2}(c+\bar{c}).
\ee

  We believe this is a remarkable result in the sense that we arrived here just using spectral flow associated to the system, and going to the large coupling regime.

\bigskip

\section{Discussions and Conclusions}\label{sec5}

\subsection{Quick summary}
We believe we have made a rather important and general discovery in this paper. We have shown that at the very edges of parameter space of a relativistic quantum field theory, there are emergence of non-Lorentzian symmetry. After having laid out the reasons why we believe this to be true in general in the introduction, we have focussed on a theory of free massless scalars in two dimensions and shown that for current-current deformations on this theory. For generic values of the deformation parameter, the theory of course remains conformal but at the very end of the flow, the spacetime algebra changes from the relativistic to the Carrollian conformal algebra. We dealt with two different types of current-current deformations, the symmetric and the antisymmetric, and carefully considered both the classical and quantum versions of the flow. There were some very interesting differences between them, indicating that the flow proceeds in very different routes, but both end up with Carrollian conformal symmetries at the end of the flow. The differences were particularly visible when we considered the quantum mechanical deformed theories. The symmetric deformation proceeded through a continuous change of the vacuum state, dictated by a Bogoliubov transformation, which the anti-symmetric deformation manifested itself as a change of boundary condition leading to an interpretation in terms of spectral flow. We summarized our findings at the beginning in Sec.\eqref{sec2} and in particular in Table \eqref{tab1}. That a spectral flow is able to generate a change in the spacetime algebra was a particularly intriguing outcome of our analysis. 

Our analysis in the paper raises the intriguing possibility that renormalisation group flows as pioneered by Wilson \cite{RevModPhys.47.773} may have its limitations and may need re-examination as there are intriguing features that appear in relativistic QFTs when they are deformed to either very low or very high energies. 

\subsection{Discussions}
There are numerous questions, confusions and a very large number of directions of open research emerging from our present work. We discuss some of them below. 
\begin{itemize}
\item[$\square$]{\em{The Galilean limit.}} In our introduction, we led with the observation that at very low energies, the effective description of a relativistic QFT should be Galilean. However, we have offered precious little in terms of concretising this point. It turns out that the Galilean limit is more subtle than the Carrollian one (see Appendix \ref{appC}) and the places where the symmetry emerges may be given by complex values of the deformation parameter. The $d=2$ problem is further complicated by the fact that the Carrollian and Galilean algebras are isomorphic leading to issues with identification of ultra-high v/s ultra-low energies. We are currently investigating this and hope to report on it in the near future when some of the associated subtlety is suitably addressed. 

\item[$\square$]{\em{Other QFTs.}}  We believe the story we are trying to tell is a generic one, i.e. there is the emergence of Galilean symmetries at very low energies and Carrollian symmetries at very high energies for {\em{any relativistic quantum field theory}}. So, we need to move beyond our present example and repeat our analysis for other dimensions and other QFTs or understand a general scheme which allows us make this argument. 

\item[$\square$]{\em{Other kind of deformations.}} Even within the realms of the 2d scalar theory, one needs to understand the emergence of Carroll in the extreme UV by considering irrelevant deformations. This is of course a very difficult problem, but the recent progress in integrable deformations like the $T\bar{T}$ deformations can be studied carefully and there are indications that Carroll structures would emerge even there. In this case, reinterpreting the current-current deformations as deformations of the worldsheet theory would be useful. 

\item[$\square$]{$\sqrt{T\overline{T}}$ \em{deformation}.} Recently in \cite{Rodriguez:2021tcz}, it was argued that a $\sqrt{T\overline{T}}$ operator added to a 2d CFT changed the symmetries to BMS$_3$ and this was not to be thought of in terms of any flow, but a non-linear map.  Later, following \cite{Bagchi:2022nvj},  it was realised \cite{Tempo:2022ndz, Parekh:2023xms} that  these are indeed interpretable as flows. This $\sqrt{T\overline{T}}$ type deformation is different \cite{Ferko:2022cix} from $T\bar{T}$ like irrelevant deformations, since here the stress tensors are just those of the undeformed seed theory, and do not generate iteratively defined flows. Such a theory with a square-root operator is hard to quantize. However, from the understanding that $\sqrt{T\overline{T}}$ can be written as a current-current bilinear operator, our way of looking into flows triggered by marginal deformations provides insight into the quantum nature of these square-root deformations. The relation between $\sqrt{T\overline{T}}$ and the current bilinears at the quantum level of course depends on operator ordering and this needs to be carefully treated.

\item[$\square$]{\em{Holography.}} Carrollian symmetries are generically associated with null manifolds. Two of the most important ones are the horizons of black holes and the null boundary of asymptotically flat spacetimes. The deformation of the relativistic 2d scalar theory can be taken as an example of something generic happening to any 2d CFT. In the context of AdS/CFT, if we think of the deformation taking the theory on the boundary and moving it into the bulk, the Carroll CFTs that emerge should be the end points of these flows. The very deep bulk region of AdS can be thought of as flat space where the effects of the AdS radius cannot be felt. From the point of view of symmetries, this can be thought of as a change of symmetries from $SO(d,2)$ to $ISO(d,1)$. This is precisely the global subgroup of the conformal Carroll algebra. For asymptotically AdS spacetimes with black holes, the flow can be thought to end on the horizon of the black hole in the bulk. The symmetries there again would be Carrollian \cite{Donnay:2019jiz}. 

\item [$\square$]{\em{c-theorem.}} We are advocating that the degrees of freedom of a relativistic QFT should decrease at very high energies, where the slow moving modes can be disregarded. This seems at complete odds with the c-theorem \cite{Zamolodchikov:1986gt,Cardy:1996xt}. The answer possibly is that the transition from relativistic to Carroll symmetries, which is the restriction of a relativistic QFT to its high energy subsector, breaks inherent assumptions of the c-theorem, like Poincare invariance and unitarity. We would like to examine this in more detail going forward. 

\item[$\square$]{\em{Renormalisation group flows in Carrollian theories: The ``top down'' approach.}} In the intriguing new work \cite{Banerjee:2023jpi}, a RG procedure customized to suit intrinsic Carrollian scalar field theories has been examined. For Carroll theories, energy scales and spatial scales are separated in contradiction to Wilsonian intuition, leading to drastic changes in the notion of canonical mass dimensions of operators. As a consequence, a spatial derivative added to a purely Carrollian electric scalar ($\cL\sim \dot\phi^2$) becomes relevant as we move towards the IR, ultimately reinstating Lorentz invariance in the process. One can think of this as a complimentary version of the picture presented here, where we have marginal deformations of scale invariant scalars triggering flows which eventually kill Lorentz invariance, and leads to a Carrollian electric scalar theory. It would be of fundamental importance to understand this dual interplay between local and ultra-local theories while going forward. 

\item[$\square$]{\em{Re-interpreting Wilson.}} Wilson's original formulation of RG flows \cite{RevModPhys.47.773} was in terms of ``slow" and ``fast" modes, where one integrated over the fast modes in order to get an effective theory of slow modes valid in the lower energy regime. We have just mentioned that in the intrinsically Carrollian theory, the notions of operator dimension and hence RG flows change drastically \cite{Banerjee:2023jpi}. The reduction of degrees of freedom of the relativistic theory, as it enters a very high energy domain where only modes traveling at or near lightspeed are of importance, might be an observer dependent redistribution of what are ``slow" modes and ``fast" modes to this high energy observer. Perhaps the Wilsonian paradigm can still be made to work, but the observer dependence has to be carefully reconsidered{\footnote{We thank Shahin Sheikh-Jabbari for illuminating conversations on this point and stressing this to us.}}.

\end{itemize}

It is clear that our current work has raised numerous interesting issues and indicated new directions of research. We hope to be back with more details on the points discussed above as well as other related problems very soon.

\bigskip

\section*{Acknowledgements}
The authors would like to thank Rudranil Basu, Joydeep Chakrabortty, Ritankar Chatterjee, Stephane Detourney, Priyadarshini Pandit and especially Shahin Sheikh-Jabbari for many fruitful discussions. 

\medskip

\noindent AB is partially supported by a Swarnajayanti Fellowship of the Department of Science and Technology and Science and Engineering Research Board (SERB) and also by the following SERB grants SB/SJF/2019-20/08, CRG/2022/006165. ArB is supported in part by an OPERA grant and a seed grant NFSG/PIL/2023/P3816 from BITS-Pilani. ArB would like to thank Kyoto University, TU Wien, CNRS Paris  and IIT Kanpur for kind hospitality during the various stages of this work. SM is supported by grant number 09/092(1039)/2019-EMR-I from Council of Scientific and Industrial Research (CSIR). DM is supported by a Young Scientist Training Program (YST) Fellowship from Asia Pacific Center for Theoretical Physics (APCTP). The work of HM is supported by the Institute for Basic Science (IBS-R003-D1).

\newpage

\appendix
\section*{APPENDICES}
\section{Computations of Poisson brackets}
\subsection*{a) Sugawara construction for stress tensors}
To touch base, let us explicitly show how we arrive at the algebra of undeformed Sugawara stress tensors, in this case, for the anti-holomorphic one:
\bea
	\left\{ \overline{T}(\sigma),  \overline{T}\left(\sigma^{\prime}\right)\right\}_{\mathrm{PB}} 
	&=\frac14\sum_{i,j} \left\{ \overline{{J}}_i^2(\sigma),\overline{{J}}_j^2(\sigma')\right\}_{\mathrm{PB}} \nn
	&=\sum_{i,j} \left\{ \overline{{J}}_i(\sigma),\overline{{J}}_j(\sigma')\right\}_{\mathrm{PB}} 
	\overline{{J}}_i(\sigma)\overline{{J}}_j(\sigma')\nn
	&=-\sum_{i} \left[\pd_{\sigma}\delta(\sigma-\sigma')\right]
	\overline{{J}}_i(\sigma)\overline{{J}}_i(\sigma')\nn
	&=-\sum_{i} \left[\pd_{\sigma}\delta(\sigma-\sigma')\right]
	\overline{{J}}_i(\sigma)\left\{\overline{{J}}_i(\sigma)
	-(\sigma-\sigma')\pd_\sigma\overline{{J}}_i(\sigma)+\mathcal{O}((\sigma-\sigma')^2)\right\}\nn
	&=-2  \overline{T}(\sigma) \partial_\sigma \delta\left(\sigma-\sigma^{\prime}\right)
	-\left(\partial_\sigma  \overline{T}(\sigma)\right) \delta\left(\sigma-\sigma^{\prime}\right)
\ena

\subsection*{b) Emergent BMS}
We show here the Poisson bracket computation for the deformed generators as described in \eqref{contract}. We only focus on one bracket and others can follow:
\bea
	&\left\{L^{(g)}(\sigma),L^{(g)}(\sigma')\right\}_{\rm PB}\nn
	&=\left\{(\Pi_+{\chi}'_++\Pi_-{\chi}'_-)(\sigma),(\Pi_+{\chi}'_++\Pi_-{\chi}'_-)(\sigma')\right\}_{\rm PB}\nn
	&=\Pi_+(\sigma){\chi}'_+(\sigma')\left\{{\chi}'_+(\sigma),\Pi_+(\sigma')\right\}_{\rm PB}
	+{\chi}'_+(\sigma)\Pi_+(\sigma')\left\{\Pi_+(\sigma),{\chi}'_+(\sigma')\right\}_{\rm PB}
	+( {\,}_+\, \to\,  {\,}_-)\nn
	&=\left\{\Pi_+(\sigma){\chi}'_+(\sigma')
	+{\chi}'_+(\sigma)\Pi_+(\sigma')\right\}
	\pd_\sigma\delta(\sigma-\sigma')
	+( {\,}_+\, \to\,  {\,}_-)\nn
	&=\left\{2\Pi_+(\sigma){\chi}'_+(\sigma)-(\sigma-\sigma')\Pi_+(\sigma){\chi}''_+(\sigma)
	-(\sigma-\sigma'){\chi}'_+(\sigma)\Pi'_+(\sigma)\right\}\pd_\sigma\delta(\sigma-\sigma')
	+( {\,}_+\, \to\,  {\,}_-)\nn
	&=2\Pi_+{\chi}'_+\pd_\sigma\delta(\sigma-\sigma')+(\Pi_+{\chi}'_+)'\delta(\sigma-\sigma')+( {\,}_+\, \to\,  {\,}_-)\nn
	&=2L^{(g)}(\sigma)\pd_\sigma\delta(\sigma-\sigma')+\left(\pd_\sigma L^{(g)}(\sigma)\right) \delta(\sigma-\sigma')
\ena
In both of the above cases we have used the formula for simplification
\be
	(x-a)^{n+1} \pd_x \delta(x-a)=-\delta_{n,0}\,\delta(x-a).
\ee
Rest of our poisson brackets can be calculated likewise.


\section{Going Galilean with $J\overline{J}$}\label{appC}
As we discussed in the main text, going to the Galilean theory using a $J\bar{J}$ deformation is much more subtle. Remember that classically \eqref{waveeqndef} makes sure the action goes to a Galilean boost invariant one when $\a = -1$ and the characteristic speed of light goes to infinity. In the quantum theory, things are not so quick since in this case, there is an inherent flip between space and time directions as compared to Carroll. Hence one has to consider generators and their contractions accordingly. The most important ingredient in this story is an automorphism at the level of Virasoro algebra, viz, $\cL_n \to \cL_n$ and $\overline{\cL}_n \to -\overline{\cL}_{-n}$ which keeps the commutation relations \eqref{Vir1} unchanged. Starting with this definition one can build what is known as the ``flipped representation" of the Virasoro algebra \cite{Bagchi:2020fpr}. At the level of stress tensors \eqref{Lmodes}, this results in $\overline{T}(\sigma) = -\overline{T}(\sigma)$ and can be equivalently taken care of by replacing:
\be\label{flippity}
\bar{j}_n \to i\bar{j}_{-n}
\ee
 in the Sugawara construction. Note that under this automorphism, the nature of the erstwhile relativistic highest-weight vacuum changes to a hybrid of highest-weight-lowest-weight case. 
\medskip

Following through with the addition of the $J\bar{J}$ term and the computation of deformed stress tensors, one can find the generators of the symmetry algebra in this case reads:
\begin{eqnarray}\label{NRgen}
\mathcal{J}_n(\a)^{NR} = \mathcal{L}_n(\a) + \bar{\mathcal{L}}_{n}(\a), \quad \mathcal{H}_n(\a)^{NR} = \frac{1}{(\cosh^2\theta + \sinh^2\theta)}\bigg(\mathcal{L}_n(\a) - \bar{\mathcal{L}}_{n}(\a)\bigg).
\end{eqnarray}
Note that these definitions flip between the canonical Hamiltonian and angular momentum generators, a fact which we alluded to in the main text. Also note these
generators do not contain any mixing of modes unlike the ones in \eqref{GHdef}. This stems from the fact that when we crank up the coupling of the $J\bar{J}$ term, the vacuum structure does not change as \eqref{bogodef} does not imply a Bogoliubov transformation owing to \eqref{flippity}. However, the above generators still lead us to the interpolating algebra in \eqref{interpol}, which still degenerate to BMS$_3$ at the same point as in the Carrollian case. This is expected, since as we mentioned before, in two dimensions, the algebras of Carrollian and Galilean conformal group are isomorphic, and these above generators \eqref{NRgen} then just become well known Galilean Conformal Field Theory (GCFT) generators \cite{Bagchi:2009my}. 
\section{String action in the presence of B-field}\label{Bfield}

The string sigma model in a general background with a NS-NS flux $(G,B)$ is described by
\be
	S=\frac12\int d^2\sigma\left(\sqrt{-h}\,h^{ab}\pd_a X^\mu \pd_b X^\nu G_{\mu\nu}
	+\epsilon^{ab}\pd_a X^\mu \pd_b X^\nu B_{\mu\nu}\right)
\ee
where the string tension $\a'$ is appropriately normalized. Let's choose a conformal gauge for the worldsheet
\be
	h_{ab}=\begin{pmatrix} e^{2\w} & 0 \\ 0 & -e^{2\w} \end{pmatrix}
\ee
and noting $\e^{01}=-\e^{10}=1$, then
\be
	S_{\rm c.g.}=\int d^2\sigma\left(\frac12\dot{X}^\mu \dot{X}^\nu G_{\mu\nu}
	-\frac12{X'}^\mu {X'}^\nu G_{\mu\nu}
	+\dot{X}^\mu {X'}^\nu B_{\mu\nu}\right).
\ee
Assuming the background field $B$ to be a constant, the conjugate momenta are
\be
	\Pi_\mu = \frac{\delta S_{c.g.}}{\delta \dot{X}^\mu}
	= \dot{X}^\nu G_{\mu\nu} + {X'}^\nu B_{\mu\nu}
\ee
which leads to:
\be
	\dot{X}^\mu=G^{\mu\rho}\Pi_\rho-G^{\mu\rho}B_{\rho\sigma} {X'}^\sigma
\ee
The associated Hamiltonian density is obtained by performing the Legendre transform
\be
	\cH=\Pi_\mu \dot{X}^\mu-\cL
	=\frac12\dot{X}^\mu \dot{X}^\nu G_{\mu\nu}
	+\frac12{X'}^\mu {X'}^\nu G_{\mu\nu}
\ee
which turns out to be only implicitly dependent on $B$.
Re-expressing the Hamiltonian density in terms of a set of canonical variables,
\bea
	\cH^{(G,B)}
	&=\frac12G^{\mu\nu}\Pi_\mu \Pi_\nu- \Pi_\mu G^{\mu\rho}B_{\rho\nu} {X'}^\nu
	+\frac12\left(-B_{\sigma\nu}G^{\nu\a}B_{\a\rho}\right){X'}^\rho{X'}^\sigma
	+\frac12{X'}^\mu {X'}^\nu G_{\mu\nu}
\ena
For a particular setting of our particular interest: two-dimensional target space with the B-field given by
\be
	B_{\mu\nu}=\Lambda\,\e_{\mu\nu}
	\qquad{\rm with}\qquad
	\e_{12}=-\e_{21}=1
\ee
one finds
\be
	-B_{\sigma\nu}G^{\nu\a}B_{\a\rho}=\Lambda^2\,G^{-1}G_{\rho\sigma}
	\qquad\quad
	G={\rm det}\,G_{\mu\nu}
\ee
thus the Hamiltonian density can be written as:
\be
	\cH^{(G,\Lambda)}=\frac12G^{\mu\nu}\Pi_\mu \Pi_\nu- \Lambda\,\Pi_\mu G^{\mu\rho}\e_{\rho\nu} {X'}^\nu
	+\frac12{X'}^\mu {X'}^\nu G_{\mu\nu}(1+\Lambda^2\,G^{-1})
\ee
Furthermore, setting $G_{\mu\nu}=\delta_{\mu\nu}$ with
	$\delta_{\mu\nu}={\rm diag}\,(1,1)$
yields the Hamiltonian density
\be
	\cH^{(\delta,\Lambda)}=\frac12\Pi_\mu \Pi^\mu + \Lambda\,\e_{\mu\nu}{X'}^\mu\Pi^\nu
	+\frac12{X'}^\mu {X'}_\mu (1+\Lambda^2).
\ee
By rescaling the above 
the dynamical fields $X\to X/\sqrt{1+\Lambda^2}$ while $\Pi \to \sqrt{1+\Lambda^2}\,\Pi$,
and replacing $\Lambda\to -g$ as well as redefining $X^\mu=X_\mu \to \phi_i$, we see
\be
	\cH^{(g)}=\frac{1}{2}\sum_{i=1,2}\left\{(1+g^2)(\Pi_i)^2 +({\phi}'_i)^2\right\}
	+ g\,\left(\Pi_1{\phi}'_2-\Pi_2{\phi}'_1\right)
\ee
which coincides with our current-current deformed case in \eqref{hamil.in.phi12}. A detailed discussion of the $g\to \infty$ case for the above, i.e. that of a null string with a Kalb-Ramond field will appear soon in \cite{upcoming}.

\section{Hamiltonian structure of general deformations}\label{appD}

A general current-current deformation to the multiple scalar field action, when put into the relevant Hamiltonian form, looks like:

\be{}
\mathcal{H} = A_{ij}\pi^i\pi^j+B_{ij}\phi'^i\phi'^j+C_{ij}\pi^i\phi'^j.
\ee
Where $\phi^i$ and $\pi^i$ are canonically conjugate variables in the undeformed system and $A,B,C$ are functions of a deformation parameter $\lambda$, say. We also have
\be{}
\lim_{\lambda\to 0}A_{ij} = \delta_{ij},~~\lim_{\lambda\to 0}B_{ij} = \delta_{ij},~~\lim_{\lambda\to 0}C_{ij} = 0.
\ee
Note that even with a general current-current deformation, the Hamiltonian remains quadratic in conjugate variables $\pi^i$ and $\phi'^i$ since only these quadratics have the dimension of energy density. The question we ask at this point: \textit{What is the criterion that the above Hamiltonian will reduce to a Carrollian dynamical system?}. Or put into a different form, when would we have,
\be{}
\{\mathcal{H}(x),\mathcal{H}(x')\} = 0,
\ee
which is characteristic of Carroll-invariant dynamics.  A way to address this problem would be to put the above Hamiltonian in a normal form (assuming we work with two scalar fields)
\be\label{symp}
\mathcal{H}_{d}  = a\Pi^2_{+}+b\Pi^2_{-}+c\phi'^2_{+}+d\phi'^2_{-},
\ee
such that $\Pi_{+},\phi'_{+}$ and $\Pi_{-},\phi'_{-}$ are canonically conjugate set of variables. Now we demand that in some asymptotic limit, the eigenvalues $a,b,c,d$, having quadratic conjugate structure amongst them, will either pairwise vanish or pairwise diverge, so that we are only left with (at most) two out of the four dynamical variables. Now consider the cases:
\\ \\
\textbf{Case I:} a and b (or c and d) survive  $\rightarrow$ Electric (Magnetic) Carroll dynamics.\\
\textbf{Case II:} a and d (or b and c) survive$\rightarrow$ Mix of Electric/Magnetic Carroll dynamics.\\
\textbf{Case III:} a and c (or b and d) survive $\rightarrow$  A single scalar field.
\\ \\
For example, consider the asymptotic structure of a generic Hamiltonian density concerning two scalar fields:
\be\label{hdensity.lesson}
	\cH=\frac{\cN_1}{2}\left\{\e_1\Pi_1^2+\frac{1}{\e_1}(\phi'_1)^2\right\}
	+\frac{\cN_2}{2}\left\{\e_2\Pi_2^2+\frac{1}{\e_2}(\phi'_2)^2\right\}.
\ee
Clearly, depending on the eigenvalues of the general Hessian matrix and the asymptotic behaviour of $\cN_i$ and $\epsilon_i$, we can reach either of the three cases mentioned above. This logic is readily extended to higher number of scalar fields as well. For this to work, one needs to be able to symplectically diagonalize Hamiltonians of the form \eqref{symp} in terms of canonically conjugate variables. In classical mechanics, such a symplectic diagonalization is possible in case the criterion of Williamson's Theorem \cite{Arnold:1989who} are met, but an extension to field theories seem absent from the literature. 

\section{Spectral flow for WZW currents}\label{appwzw}
Let us review the central results pertaining to the spectral flow of $SL(2,\mathbb{R})$ WZW theory following \cite{Maldacena:2000hw,McElgin:2015eho,Hemming:2004gd} to better appreciate the results in Sec.\eqref{sec4}. The currents in the WZW model can be expressed in terms of the components $t^a$ of the Lie algebra as
\begin{equation}
	J=J^at^a \quad \text{and} \quad J^a=\text{Tr}(t^a J)\ ,
\end{equation}
where the currents are explicitly given by
\begin{equation}
	J(z)=-\frac{k}{2}\partial_z g g^{-1} \quad \text{and} \quad \bar{J}(\bar{z})=\frac{k}{2}g^{-1}\partial_{\bar{z}} g \ .
\end{equation}
$g$ is a group element of the Lie group which in our case is $SL(2, \mathbb{R})$. Choosing the specific parametrization of the group element $g$\footnote{This parametrization helps in recovering the BTZ geometry from the WZW action outside the black hole.}
\begin{equation}
	g=e^{u \sigma_3}e^{\rho \sigma_1}e^{-v \sigma_3}\ ,
\end{equation}	
and choosing a convenient basis for $SL(2,\mathbb{R})$ as,
\begin{equation}
	t^+=t^0+t^1=\frac{1}{2}(-\sigma^2 +i \sigma^1)\ ,\ t^-=t^0-t^1=\frac{1}{2}(-\sigma^2 -i \sigma^1)\ ,\ t^3=\frac{i}{2}\sigma^3\ ,
\end{equation}
we see that the current follows the algebra:
\begin{eqnarray}
	[J^3_n,J^{\pm}_m]&=&\pm iJ^{\pm}_{n+m}\ ,\\
	{[J^+_n,J^-_m]}&=&-2iJ^3_{n+m}-kn\delta_{n+m,0}\ ,\\
	\label{eq:sl2R-abelian}
	{[J^3_n, J^3_m]}&=&\frac{k}{2}n\delta_{n+m,0}\ .
\end{eqnarray}
In the above set of definitions, the $\sigma^i$'s are the Pauli matrices and $u,\rho$ and $v$ can be interpreted as target space coordinates. Note that the \emph{abelian sector} of the above algebra generated by $J^3_n$-s follows an \emph{abelian Kac-Moody algebra}.
The chiral symmetry of the WZW model now enables us to generate new solutions as
\begin{equation}
	g(x^+,x^-) \rightarrow e^{-iw_+x^+ t^3}g(x^+,x^-)e^{iw_-x^- t^3}\ ,
\end{equation}
where $x^{\pm}=\tau \pm \sigma$ and $w_{\pm}$ parametrize the \emph{spectral flow}. Under this spectral flow, the \emph{abelian} component of the currents\footnote{There are of course two non-abelian components which transform under spectral flow as $J^{\pm} \rightarrow \tilde{J}^{\pm} \equiv J^{\pm} e^{\mp w_{+}x^+}$ and $\bar{J}^{\pm} \rightarrow \tilde{\bar{J}}^{\pm} \equiv \bar{J}^{\pm} e^{\mp w_{-}x^-}$ while their modes transform as $J_n^{\pm} \rightarrow J^{\pm}_{n \pm i w_+}$ and  $\bar{J}_n^{\pm} \rightarrow \bar{J}^{\pm}_{n \pm i w_-}$ respectively. However, we for the purpose of our current work, we focus on the sector following an \emph{abelian} Kac-Moody algebra.} transform as
\begin{eqnarray}
	\label{eq:Jspectralflowsl2R}
	J^3 \rightarrow \tilde{J}^3 \equiv J^3 + \frac{k}{2}w_{+}\ &,&\ \bar{J}^3 \rightarrow \tilde{\bar{J}}^3 \equiv \bar{J}^3 - \frac{k}{2}w_{-}\ .
\end{eqnarray}
The transformation of the corresponding modes are given by
\begin{eqnarray}
	J^3_n \rightarrow J^3_n + \frac{k}{2}w_{+}\delta_{n,0}\ &,&\ \bar{J}^3_n \rightarrow \bar{J}^3_n - \frac{k}{2}w_{-} \delta_{n,0}\ .
\end{eqnarray}
Using the usual Sugawara construction, we can further define the stress tensor as
\begin{equation}
	T=\frac{1}{k}\eta_{ab} J^a J^b=\frac{1}{2k}(J^+J^-+J^-J^- -2J^3)\ .
\end{equation} 
	Under spectral flow the above transforms as
	\begin{equation}
		\label{eq:Tspectralflowsl2R}
		T \rightarrow T+w_{+}J^3+\frac{k}{4}w_+^2\quad , \quad \bar{T} \rightarrow \bar{T}-w_{-}\bar{J}^3+\frac{k}{4}w_{-}^2\ .
	\end{equation}
Finally, the transformation rule for the Virasoro generators are given by
\begin{equation}
	L_n \rightarrow L_n +w_+ J^3_n +\frac{k}{4}w_+^2 \delta_{n,0}\quad , \quad \bar{L}_n \rightarrow \bar{L}_n -w_{-}\bar{J}^3_n +\frac{k}{4}w_{-}^2 \delta_{n,0}\ .
\end{equation}
Now one can explicitly compare between the above and \eqref{flowofL}, and see in the case of $J^1\wedge\bar{J}^2$ deformation, the spectral flow parameter is proportional to the KK momenta instead of the winding modes. However the proportionality constant involves the coupling for the $J\wedge J$ deformation which makes the marked difference, especially in the extreme limit.

\bibliographystyle{JHEP}
\bibliography{NLF}
 
\end{document}